\documentclass[a4paper,fleqn,onecolumn,usenatbib]{mnras}

\usepackage{mathptmx}
\usepackage[T1]{fontenc}
\usepackage{ae,aecompl}

\usepackage{graphicx}	% Including figure files
\usepackage{amsmath}	% Advanced maths commands
\usepackage{amssymb}	% Extra maths symbols
\usepackage{indentfirst}

\def\nn{\nonumber}

\def\be{\begin{equation}}
\def\ee{\end{equation}}
\def\beq{\begin{eqnarray}}
\def\eeq{\end{eqnarray}}
\def\nn{{\nonumber}}
\def\vp{\varphi}
\def\bga{\bar{\gamma}}
\def\bg{\bar{\Gamma}}

\renewcommand{\[}{\left[}
\renewcommand{\]}{\right]}
\title[Axion binary black hole merger]{An axion-like scalar field environment effect on binary black hole merger}

\author[Q. Yang et al.]{
Qing Yang,$^{1}$\thanks{E-mail: yangqing@bnu.edu.cn}
Li-Wei Ji,$^{2,3}$
Bin Hu,$^{1}$
Zhou-Jian Cao,$^{1}$
and Rong-Gen Cai$^{2,3}$\\
$^{1}$Department of Astronomy, Beijing Normal University, Beijing, 100875, China\\
$^{2}$CAS Key Laboratory of Theoretical Physics, Institute of Theoretical Physics,\\
Chinese Academy of Sciences, P.O. Box 2735, Beijing 100190, China \\
$^{3}$School of Physical Sciences, University of Chinese Academy of Sciences,
No.19A Yuquan Road, Beijing 100049, China}

\date{Accepted XXX. Received YYY; in original form ZZZ}

\pubyear{2017}

\begin{document}
\label{firstpage}
\pagerange{\pageref{firstpage}--\pageref{lastpage}}
\maketitle

\begin{abstract}
%This is a simple template for authors to write new MNRAS papers.
%The abstract should briefly describe the aims, methods, and main results of the paper.
%It should be a single paragraph not more than 250 words (200 words for Letters).
%No references should appear in the abstract.
Environment, such as the accretion disk, could modify the signal of the gravitational wave from the astrophysical black hole binaries. In this article, we model the matter field around the intermediate-mass binary black holes by means of an axion-like scalar field and investigate their joint evolution. In details, we consider the equal mass binary black holes surrounded by a shell of axion-like scalar field both in spherical symmetric and non-spherical symmetric cases, and with different strength of the scalar field.
Our result shows that the environmental scalar field could essentially modify the dynamics. Firstly, in the spherical symmetric case, with increasing of the scalar field strength, the number of circular orbit of the binary black hole is reduced. It means that the scalar field could significantly accelerate the merger process. Secondly, once the scalar field strength exceeds certain critical value, the scalar field could collapse into a third black hole with its mass being larger than the binary. Consequently, the new black hole collapsed from the environmental scalar field could accrete the binary promptly and the binary collides head-on between each other. In this process, there is almost no any quadrupole signal produced, namely the gravitational wave is greatly suppressed. Thirdly, when the scalar field strength is relatively smaller than the critical value, the black hole orbit could develop eccentricity through the accretion of the scalar field.
Fourthly, during the initial stage of the inspire, the gravitational attractive force from the axion-like scalar field could induce a sudden turn in the binary orbits, hence result in a transient wiggle in the gravitational waveform.
Finally, in the non-spherical case, the scalar field could gravitationally attract the binary moving toward the mass center of the scalar field and slow down the merger process.
\end{abstract}


%%%%%%%%%%%%%%%%%%%%%%%%%%%%%%%%%%%%%%%%%%%%%%%%%%

%%%%%%%%%%%%%%%%% BODY OF PAPER %%%%%%%%%%%%%%%%%%

%%++++++++++++++++++++++++++++%%
\section{Introduction}
Gravitational wave is an important prediction of Einstein's general relativity.
Its direct detection, first made by LIGO-VIRGO collaboration \cite{TheLIGOScientific:2016src,TheLIGOScientific:2016wfe,Abbott:2016nmj,TheLIGOScientific:2016pea}, not only provided a unique method to test Einstein's theory of general relativity in the strong field region, but also will open a new era in astronomy. The signals from the reported events GW150914 and GW151226 have been shown to match the waveforms predicted by general relativity for merging stellar-mass black holes \cite{TheLIGOScientific:2016src}. The corresponding initial masses of the merging black holes are $36^{+5}_{-4} M_\odot$ and $29^{+4}_{-4}M_\odot$ for GW150914 \cite{TheLIGOScientific:2016wfe,TheLIGOScientific:2016pea} and $14^{+8}_{-4}M_\odot$ and $8^{+2}_{-2}M_\odot$ for GW151226 \cite{Abbott:2016nmj,TheLIGOScientific:2016pea}. These events prove the existence of astrophysical black holes, the formation of binary block holes in nature and their consistent behavior as general relativity predicts. Besides, the source parameters estimated from comparison of numerical simulations of the gravitational-wave signal and the observations can give new insight into their astrophysical formation \cite{TheLIGOScientific:2016pea,Abbott:2016nhf}. Many more steller-mass black holes are expected to be detected by LIGO in the next few years.

A typical stellar-mass black hole with mass ranging from about $3$ to several tens of solar masses, is assumed to be formed by the gravitational collapse of a single massive star. It is inevitable at the end of the life of a star, when all stellar energy sources are exhausted. If the collapsing mass is above the Tolman-Oppenheimer-Volkoff limit, roughly $3$ solar masses \cite{1996A&A...305..871B}, a stellar-mass black hole will form. Besides, there is observational evidence for two other types of black holes, which are much more massive than the stellar-mass one. They are intermediate-mass black holes and supermassive black holes. For the former, its typical mass is ranged between $10^2$ to $10^5$ solar masses. One of the possible candidates is believed to be located in the centre of globular clusters \cite{Gebhardt:2005cy}. For the latter, it is the largest type of black holes, on the order of hundreds of thousands to billions of solar masses, and is found in the centre of almost all currently known massive galaxies \cite{Antonucci:1993sg,Urry:1995mg}.

As mentioned above, the stellar-mass black hole is formed via the collapse of the single massive star. After its formation, most of the material of the massive star have already been accreted into the black hole. Outside of the black hole, there is almost nothing. Hence, in the numerical relativity computation of the binary stellar-mass black hole merger, we can use the vacuum solution outside their event horizon. On the other hand, the intermediate-mass black holes are too massive to be formed by the collapse of a single star. There are three possible formation scenarios. The first is the merging of the stellar mass black holes or the other compact objects by means of accretion. The second is the runaway collision of massive stars in the dense stellar clusters. The third is the primordial black holes formed in the deep radiation era. In the first two channels, the black hole is likely to be surrounded by the matter materials. As for the supermassive black hole, its origin remains an open question. Astrophysical observation suggest that they are located in the center of a galaxy and can grow by accretion of matter and by merging with other black holes. All these suggest that, when simulate the merger of the intermediate-mass or supermassive black hole binaries, we need to consider the environment effect.

On the other hand, scalar field has already been introduced to study scalar radiation from BH binaries in scalar-tensor theory, triggered either by non-trivial boundary conditions \cite{Berti:2013gfa} or by non-trivial potential \cite{Healy:2011ef}. In \cite{Cao:2013osa}, GW radiation in $f(R)$ theory was studied by solving the equivalent GR equations coupled to a real scalar field. In this article, we study the binary black hole environment by modelling the matter material via an axion-like scalar field.

The rest of the paper is organized as follows. In section \ref{axion}, we will briefly introduce the axion-like scalar field.
In section \ref{large} and \ref{ms}, the evolution of the system with initially spherically distributed scalar field is considered. The former is the case with large scalar field strength and the latter are the cases with medium and small field strength. The non-spherical symmetric case is presented in section \ref{ns}. Our final conclusion is drawn in section \ref{con}. As of the mathematical equations and numerical method, they are presented in Appendix \ref{appendixA}.

In most of the paper we'll stay in the geometric units\footnote{The transformation from the natural to the geometric unit is presented in Appendix \ref{appendixB}},
in which the speed of light $c$ and the gravitational constant $G$ are normalized to $1$. So only one dimension is left, which we choose to be length. We also define a new unit $M$, which is related to meter via $1\mathrm{M}=10^{11}\mathrm{m}$. We employ the following notation: The Greek indices $(\mu,~\nu,~...)$ refer to four-dimensional space-time indices and take values from 0 to 3. The Latin indices $(i,~j,~k,~l,~...)$ refer to three-dimensional space indices and take value from 1 to 3.

\section{Axion-like scalar field}
\label{axion}
In this section, we model the matter field which surrounds the binary black hole (BBH) by means of a spherical shell composed by an axion-like scalar field. There exist several possible candidates for the extraordinary light bosons in the fundamental physics, such as the axion field for the fussy dark matter \cite{Hu:2000ke}. In this work, we consider the Hilbert-Einstein action with the axion-like scalar field written as
\beq\label{Lagrangian}
S=\int d^4x\sqrt{-g}\left[R-\frac{1}{2}g^{\mu\nu}\partial_{\mu}\phi\partial_{\nu}\phi-V(\phi)\right]
\eeq
where $V(\phi)$ is the potential of the axion-like field. In natural units it can be written as:
\beq\label{potential}
V(\phi)=m^2f^2\Big(1-\mathrm{cos}(\frac{\phi}{f})\Big)
\eeq
For our purpose of considering the interaction of axion-like scalar field and binary black holes, we will fix the mass and decay constant of the particle at $m=10^{-21}\mathrm{eV}$ and $f=0.5\mathrm{GeV}$ in the calculation, as suggested in \cite{Hui:2016ltb}.

We will consider the evolution of an equal-mass binary black hole system with a shell of axion field around it. Initially, each of the black holes has a mass parameter of 0.5M\footnote{Here we consider the intermediate-mass binary black hole, hence choose the mass of the black hole to be $0.5M\simeq 100M_\odot$.}, corresponding to an ADM mass of $0.990473\mathrm{M}$, and the separation between the two black holes is $11$M along the y-axis. For the axion shell, we will always set it to a spherical one with a radius of $120$M. Its profile along the radius direction is the following guassian type:
\beq
\phi(r)=\phi_0e^{-(r-r_0)^2/2\sigma^2}
\eeq
where $r_0$ is the center of the gaussian profile and $\sigma$ is the width parameter, we have chosen that $r_0=120\mathrm{M}$ and $\sigma=2\mathrm{M}$.

For the simpliest case, in which the whole system is spherical symmetric, we place the center of the axion shell at the mass center of the BBH system, which is just the origin of numerical domain $(x,y,z)=(0,0,0)$. we will consider the evolution of the system with three different $\phi_0$s, i.e. small, medium and large with $\phi_0=10^{-2},~10^{-3},~10^{-4}$ respectively.

To get a clear physical picture and make an estimation of how much impact the axion field will have on the dynamics of the BBH system, we can calculate the total energy of the axion field and compare it with the total mass of the black holes. The energy density of the axion field is given by:
\beq
E&:=&n_a n_b T^{ab}=\frac{1}{2}D_i\phi D^i\phi+\frac{1}{2}\dot{\phi}^2+\frac{c}{\hbar}m^2f^2[1-cos(\phi/(\sqrt{\frac{\hbar}{c}}f))]
\eeq
we can integrate the energy density of the initial profile over the whole space and make an estimation of the total energy of the axion shell. We performed this simply integration and found the total energy is $4.108\mathrm{M}$, $0.04108\mathrm{M}$ and $0.0004108\mathrm{M}$ for the large, medium and small $\phi_0$ respectively, which is about $4$, $0.04$ and $0.0004$ times the total mass of the binary black holes.

For $\phi_0=10^{-3}$ we will also consider a non-spherical symmetric case, in which we place the center of the shell at $(x,y,z)=(0,0,50)$, i.e., the shell is translated 50M along the positive direction of the x-axis, while the mass center of the two black holes is still located at the coordinate origin, and the profile of the spherical axion shell along the radius direction is not changed. So the binary is located inside the axion shell 50M away from its center.

We'll describe the evolution of these systems in the following sections.

\section{Evolutions of the System in Spherical Symmetric Large $\phi_0$ case}
\label{large}
In this section, we will present the numerical results of the evolution of the system in the spherical symmetric large $\phi_0$ case, i.e. $\phi_0$ equals to $10^{-2}$. The dynamics of the axion field and the dynamics of the binary black holes will be discussed. Besides, we will also examine the gravitational waves emitted.
%\subsubsection{Dynamics of the Axion Field}
\begin{figure}
  \centering
  \includegraphics[width=0.5\textwidth]{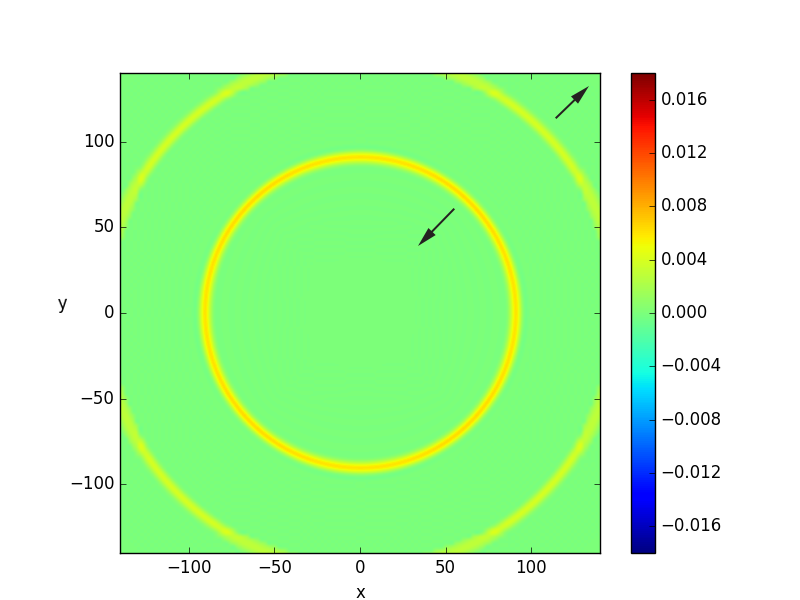}
  \caption{\small The profile of the axion field at $t=30.3\mathrm{M}$ in the large $\phi_0$ case.}\label{large_iniphi}
\end{figure}

\begin{figure}
\begin{minipage}[t]{0.3\linewidth}
\centering
\includegraphics[width=2.1in]{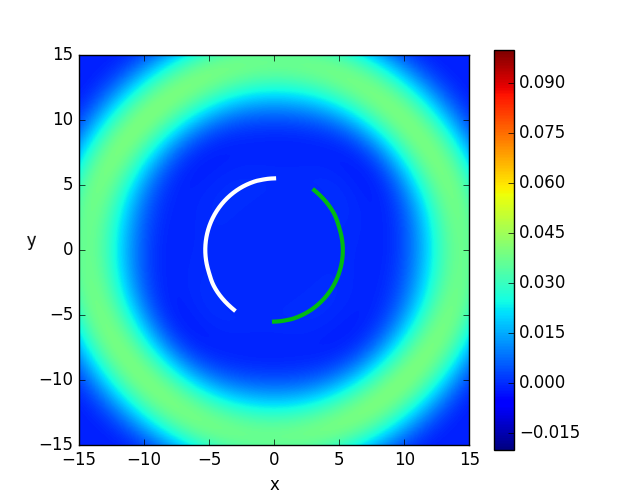}
\end{minipage}
\begin{minipage}[t]{0.3\linewidth}
\centering
\includegraphics[width=2.1in]{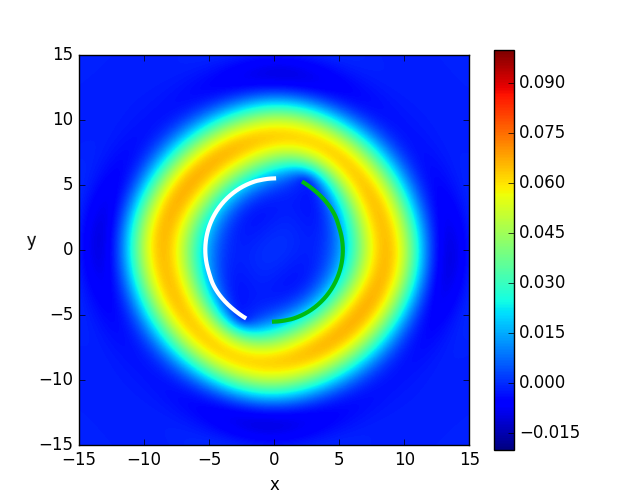}
\end{minipage}
\begin{minipage}[t]{0.3\linewidth}
\centering
\includegraphics[width=2.1in]{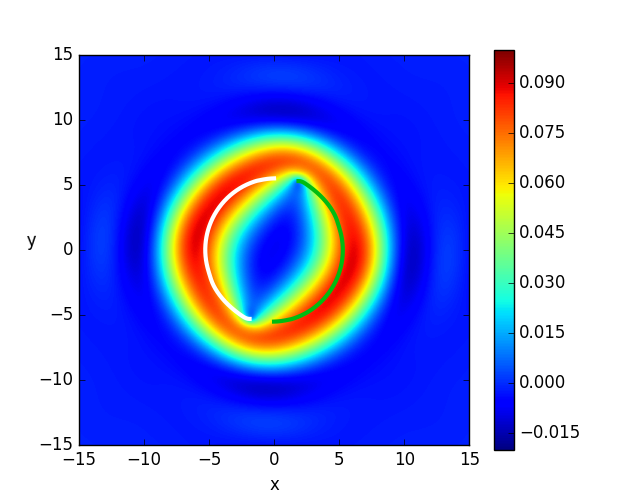}
\end{minipage}

\begin{minipage}[t]{0.3\linewidth}
\centering
\includegraphics[width=2.1in]{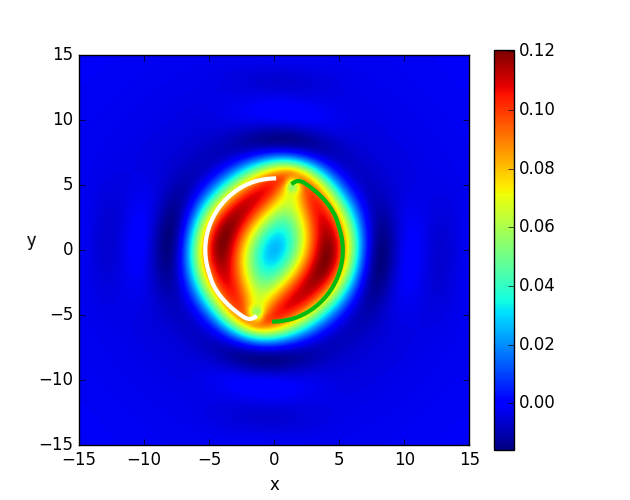}
\end{minipage}
\begin{minipage}[t]{0.3\linewidth}
\centering
\includegraphics[width=2.1in]{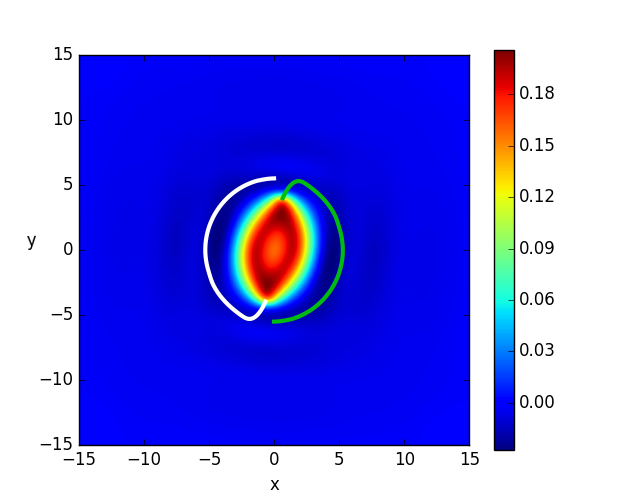}
\end{minipage}
\begin{minipage}[t]{0.3\linewidth}
\centering
\includegraphics[width=2.1in]{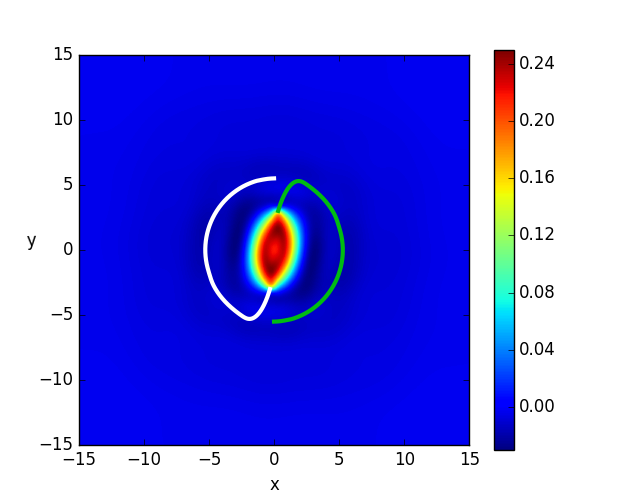}
\end{minipage}

\begin{minipage}[t]{0.3\linewidth}
\centering
\includegraphics[width=2.1in]{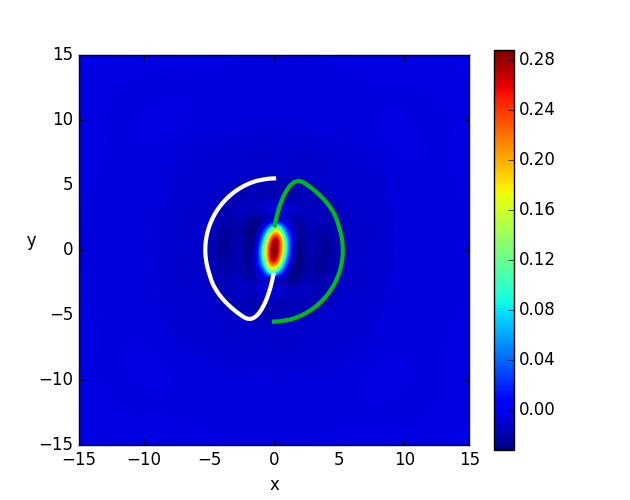}
\end{minipage}
\begin{minipage}[t]{0.3\linewidth}
\centering
\includegraphics[width=2.1in]{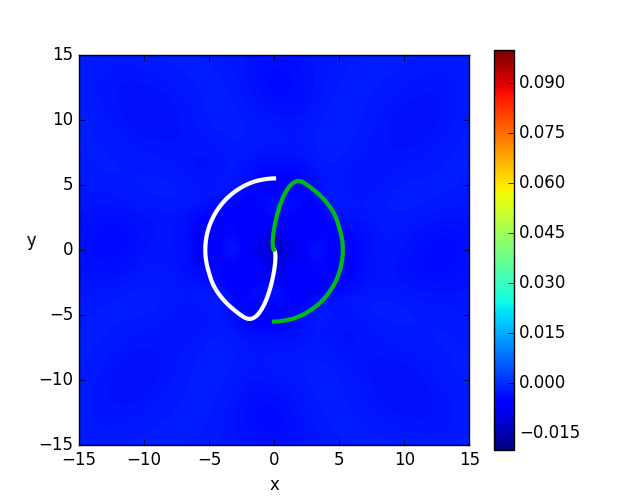}
\end{minipage}
\begin{minipage}[t]{0.3\linewidth}
\centering
\includegraphics[width=2.1in]{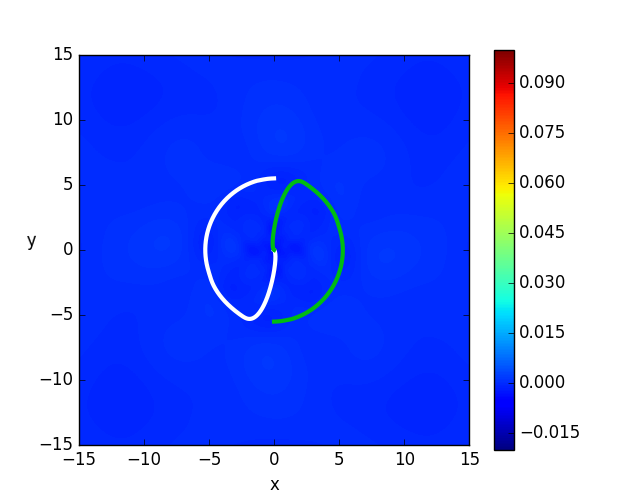}
\end{minipage}

\caption{\small The joint evolution of the scalar field and the black hole binary system in the large $\phi_0$ case is shown. The orbits of the two black holes are given in white and green lines. Different subplots represent different timesteps, which are chosen to be $t=116.15\mathrm{M},~ 126.25\mathrm{M},~131.30\mathrm{M},~
136.35\mathrm{M},~146.45\mathrm{M},~151.50\mathrm{M},~156.55\mathrm{M},~181.10\mathrm{M},~207.05\mathrm{M}$ respectively. The orbits of the black holes quickly bent to the center as soon as the axion shell passed, and turned from a semi-circle one to a head-on collision.}
\label{large_BHphi}
\end{figure}

\begin{figure}
\begin{minipage}[d]{0.48\linewidth}
\centering
\includegraphics[width=3.1in]{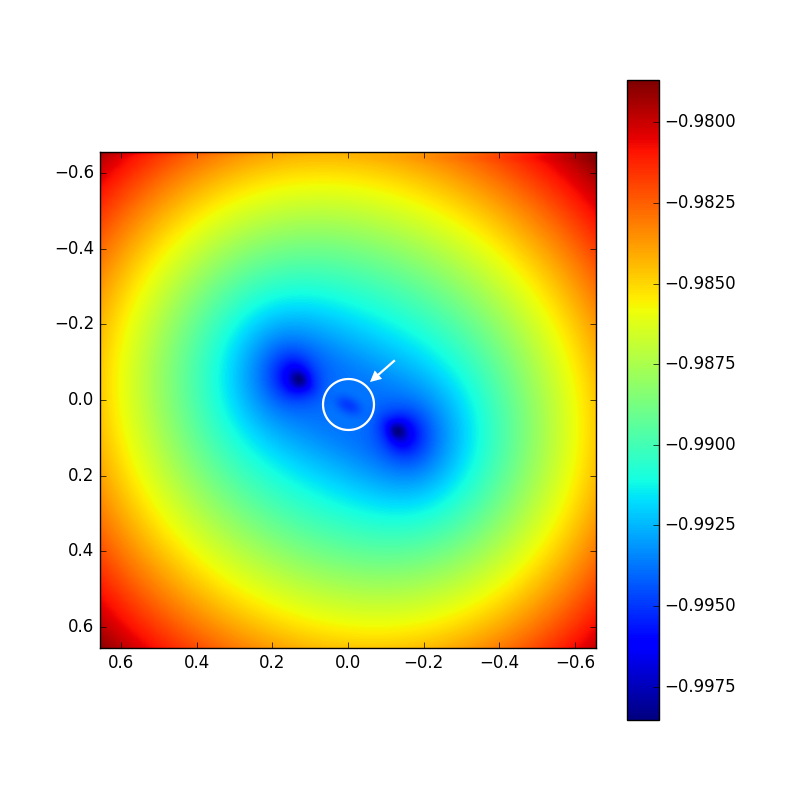}
\end{minipage}
\begin{minipage}[d]{0.4\linewidth}
\centering
\begin{minipage}[t]{0.3\linewidth}
\centering
\includegraphics[width=0.8in]{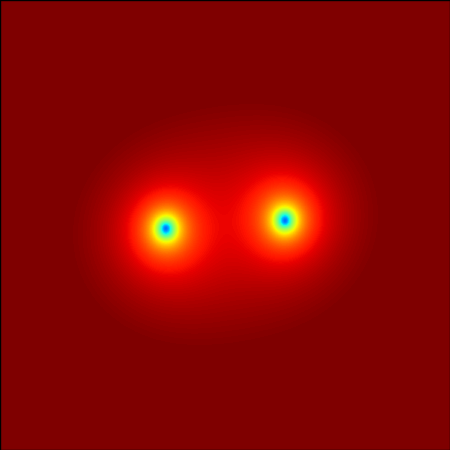}
\end{minipage}
\begin{minipage}[t]{0.3\linewidth}
\centering
\includegraphics[width=0.8in]{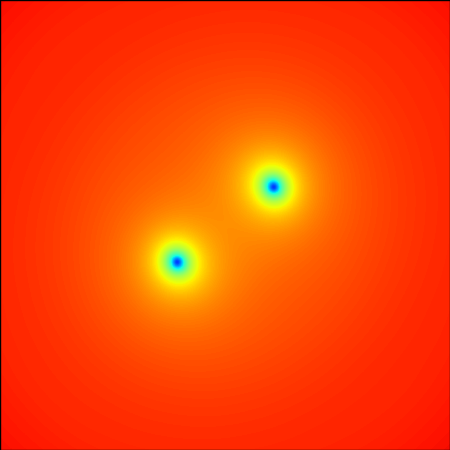}
\end{minipage}
\begin{minipage}[t]{0.3\linewidth}
\centering
\includegraphics[width=0.8in]{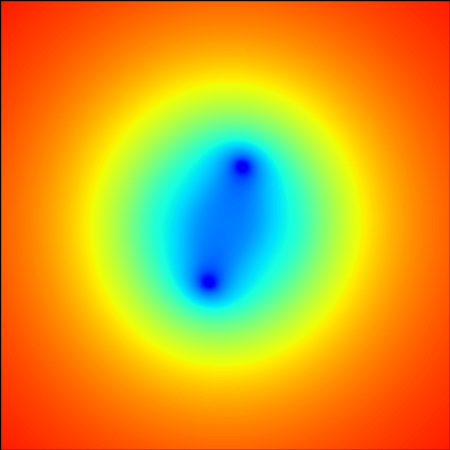}
\end{minipage}

\begin{minipage}[t]{0.3\linewidth}
\centering
\includegraphics[width=0.8in]{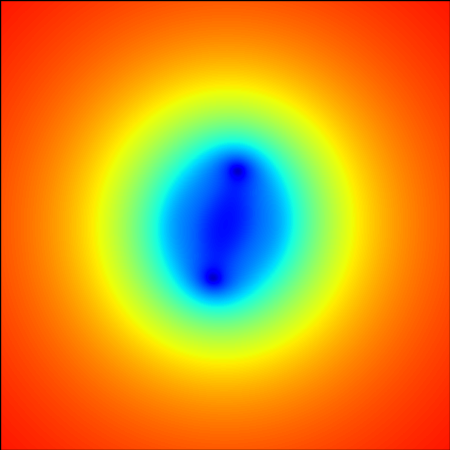}
\end{minipage}
\begin{minipage}[t]{0.3\linewidth}
\centering
\includegraphics[width=0.8in]{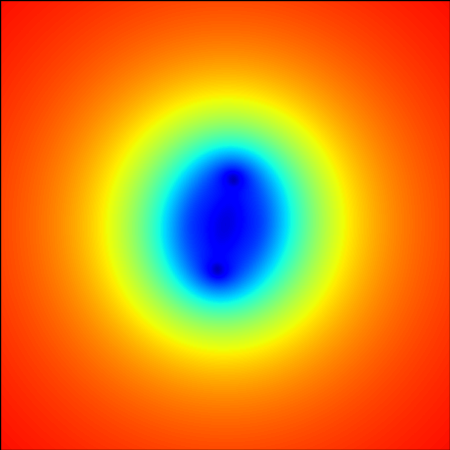}
\end{minipage}
\begin{minipage}[t]{0.3\linewidth}
\centering
\includegraphics[width=0.8in]{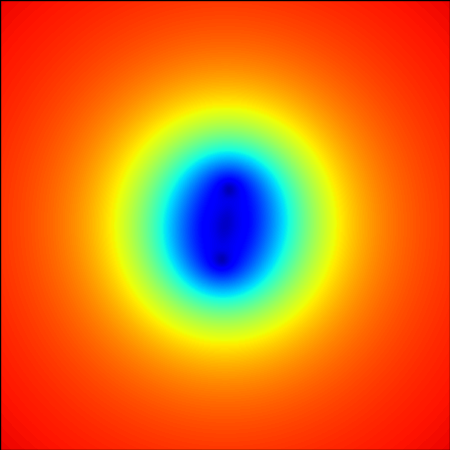}
\end{minipage}

\begin{minipage}[t]{0.3\linewidth}
\centering
\includegraphics[width=0.8in]{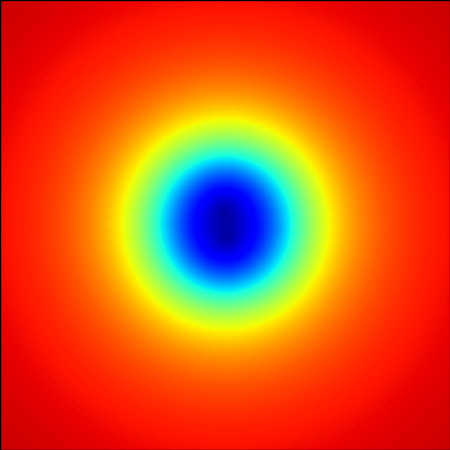}
\end{minipage}
\begin{minipage}[t]{0.3\linewidth}
\centering
\includegraphics[width=0.8in]{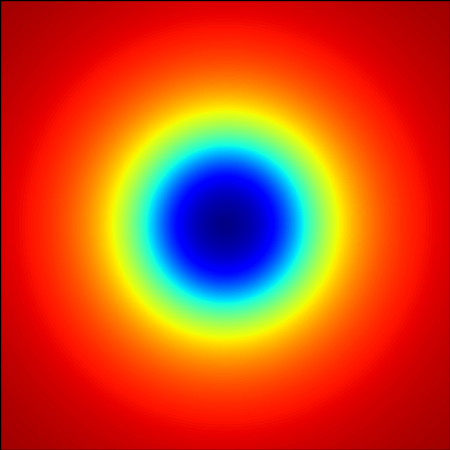}
\end{minipage}
\begin{minipage}[t]{0.3\linewidth}
\centering
\includegraphics[width=0.8in]{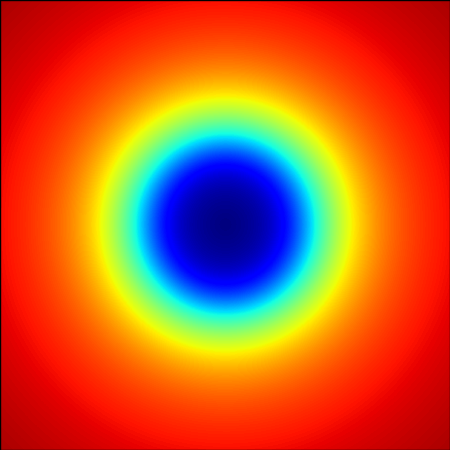}
\end{minipage}
\end{minipage}
\begin{minipage}[d]{0.1\linewidth}
\centering
\includegraphics[width=0.29in]{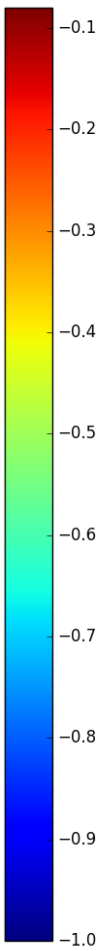}
\end{minipage}
\caption{\small $\alpha-1$ in the large $\phi_0$ case is shown, where $\alpha$ is the lapse function. Left panel: a detailed sketch of $\alpha-1$ at $t=171.70\mathrm{M}$, a third black hole can be roughly seen in the center (marked by the white circle in the figure). Right panel: sketch of $\alpha-1$ in different timesteps, which is $t=75.75\mathrm{M},~101.00\mathrm{M},~136.35\mathrm{M},~141.40\mathrm{M},~146.45\mathrm{M},
~151.50\mathrm{M},~161.60\mathrm{M},~176.75\mathrm{M},~202.00\mathrm{M}$ respectivily. The final black hole is comparable with the two initial black holes.}
\label{large_lap}
\end{figure}

We'll firstly look at the evolution of the axion shell in this case. Since the system is spherical symmetric, it is sufficient to look at the axion field in the x-y plane, so all the pictures in this section and subsequent sections will be present in the x-y plane. In Fig.\ref{large_iniphi} we show the profile of the axion field at $t=30.3\mathrm{M}$. We can see that very soon after the system starts to evolve, the gaussian shaped axion shell splits into two parts and evolves inward and outward separately. The out going part will go out of the numerical boundary in some time and will not interact with the binary black holes.

We show in Fig.\ref{large_BHphi} the evolution of the ingoing shell together with the black hole orbits. It can be seen from the plots that as the ingoing shell falls into the center, it grows denser, and when it finally impacts on the black holes, the center value has reached about 20 times the initial value. It is clear that the axion field is attracted by the BBH, and turned from spherical to an almond shape. More careful examination of the axion field shows that, after a sign change at the center, most of the scalar field is absorbed. There are no outgoing waves afterwards.

The axion field does not implode through the center. This motivated us to check if there is a third black hole formed in the center by the axion field before the two black holes collide with each other. We can check this by looking at the lapse function $\alpha$, which is related to the $tt$ component of the spacetime metric under $3+1$ decomposition:
\beq
g_{\mu\nu}dx^{\mu}dx^{\nu}=-\alpha^2dt^2+\gamma_{ij}(dx^i+\beta^idt)(dx^j+\beta^jdt)
\eeq
We can roughly say a third black hole has formed when the lapse function $\alpha$ reached zero at some point. The data of $\alpha-1$ is shown in Fig.\ref{large_lap}, so we are actually looking for $-1$ point in these figures. In the righthand sub-panel of Fig.\ref{large_lap}, we show how the lapse function evolves through different timesteps, and the lefthand side is a detailed sketch of the data at $t=171.7\mathrm{M}$. A third black hole is roughly seen in the center, and after the collision happened, a black hole that is much heavier than the original binary is formed.

\begin{figure}
\begin{minipage}[t]{0.5\linewidth}
\centering
\includegraphics[width=2.7in]{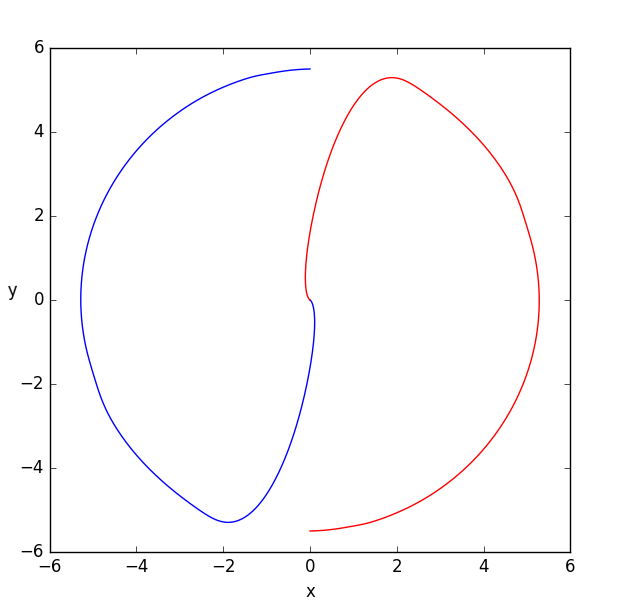}
\end{minipage}
\begin{minipage}[t]{0.5\linewidth}
\centering
\includegraphics[width=3.5in]{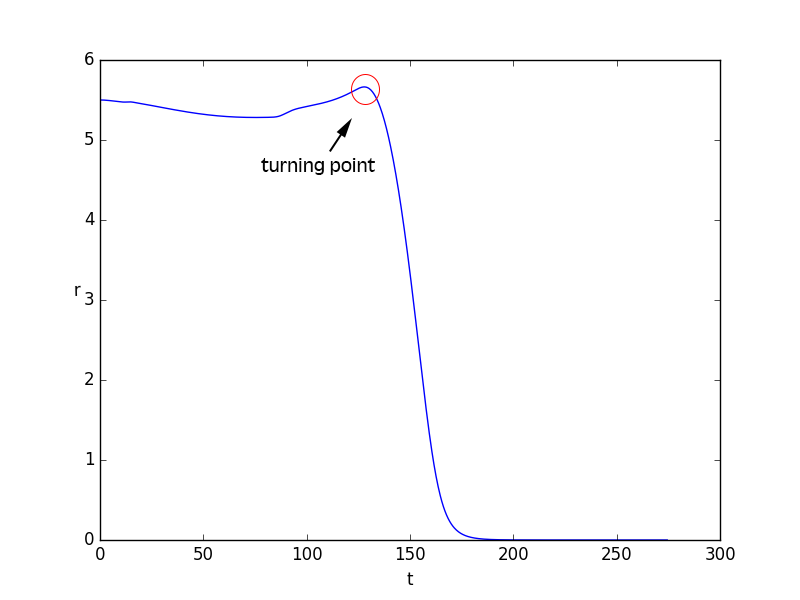}
\end{minipage}
\caption{Left panel: the orbits of the two black holes in the large $\phi_0$ case. Two different color represent two black holes. Right panel: the radial position of one of the black holes as a function of time in the large $\phi_0$ case.}\label{large_orbit}
\end{figure}

We have estimated the total energy of the axion shell at the beginning of this section, and found that for the large $\phi_0$ case, the energy of the axion field is four times the total mass of the BBH, so we expect the ingoing shell will have an obvious impact on the evolution of the black holes. In Fig.\ref{large_orbit}, we depicted the orbit of the binary black holes in the left handside panel and their radius as a function of time in the right handside panel. Consider also the joint evolution of the BBH and the axion field shown in Fig.\ref{large_BHphi}, we can see the orbit has a sharp turn when the axion field passes through, and turned from a semi-circle inspiring into a head-on collision. The radial position plot in Fig.\ref{large_orbit} also shows that the black holes are slightly attracted by the axion shell as it approaches around about $t=80\mathrm{M}$ to $t=130\mathrm{M}$, and the turning point is located at around $t=135\mathrm{M}$ when the ingoing shell starts to collide with the back holes, as is depicted in the third subplot in Fig.\ref{large_BHphi}.
The waveforms emitted in this process is plotted in Fig.\ref{GW}, we'll discuss it in the corresponding section.

\section{Evolutions of the System in Spherical Symmetric medium and small $\phi_0$ cases}
\label{ms}
\begin{figure}
\begin{minipage}[t]{0.3\linewidth}
\centering
\includegraphics[width=2.1in]{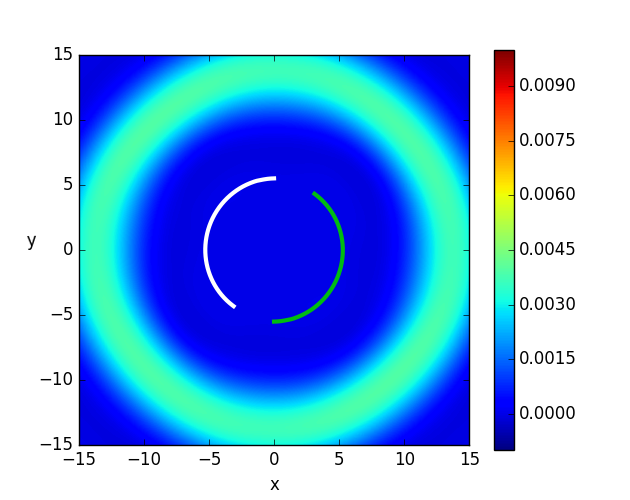}
\end{minipage}
\begin{minipage}[t]{0.3\linewidth}
\centering
\includegraphics[width=2.1in]{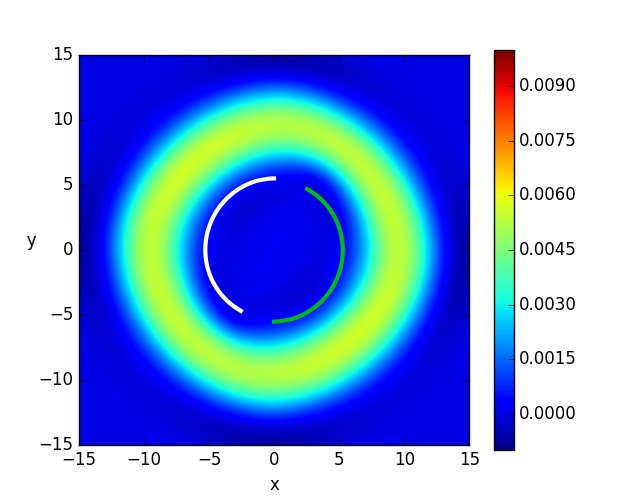}
\end{minipage}
\begin{minipage}[t]{0.3\linewidth}
\centering
\includegraphics[width=2.1in]{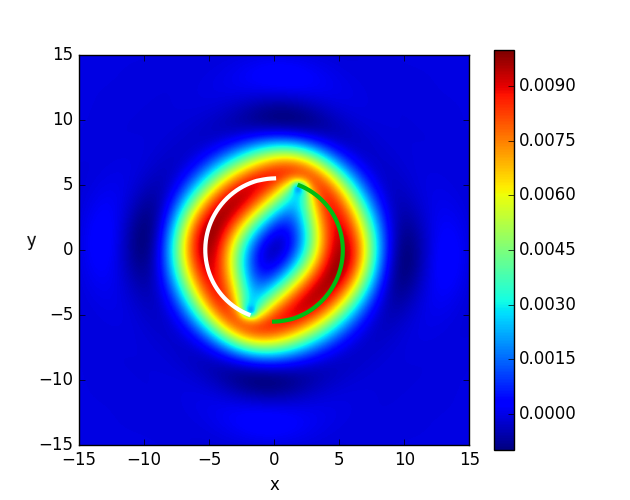}
\end{minipage}

\begin{minipage}[t]{0.3\linewidth}
\centering
\includegraphics[width=2.1in]{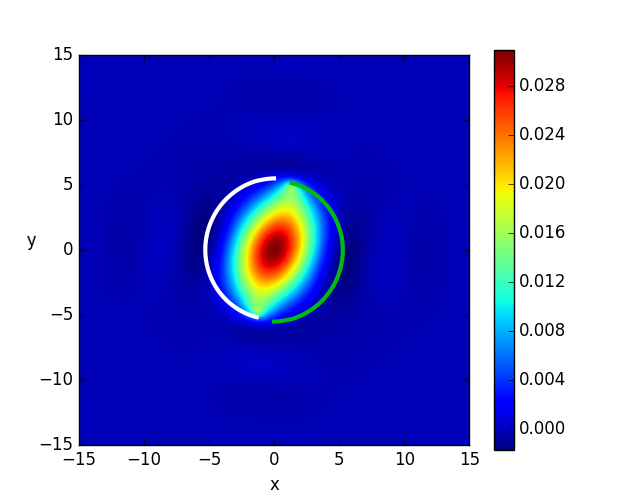}
\end{minipage}
\begin{minipage}[t]{0.3\linewidth}
\centering
\includegraphics[width=2.1in]{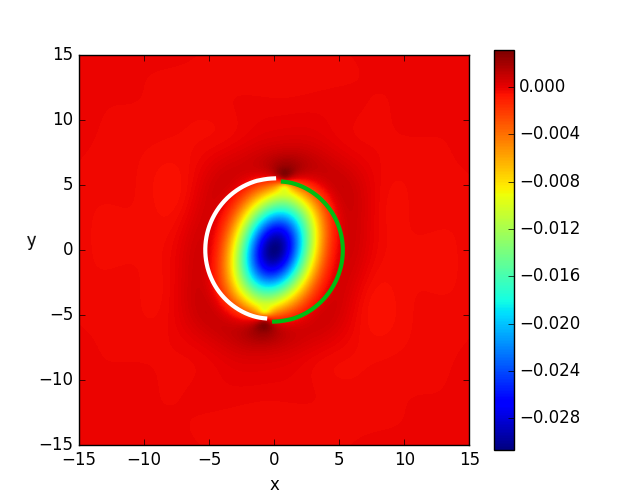}
\end{minipage}
\begin{minipage}[t]{0.3\linewidth}
\centering
\includegraphics[width=2.1in]{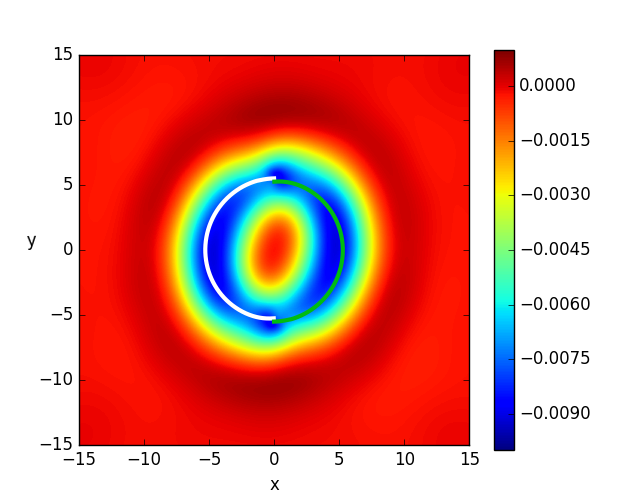}
\end{minipage}

\begin{minipage}[t]{0.3\linewidth}
\centering
\includegraphics[width=2.1in]{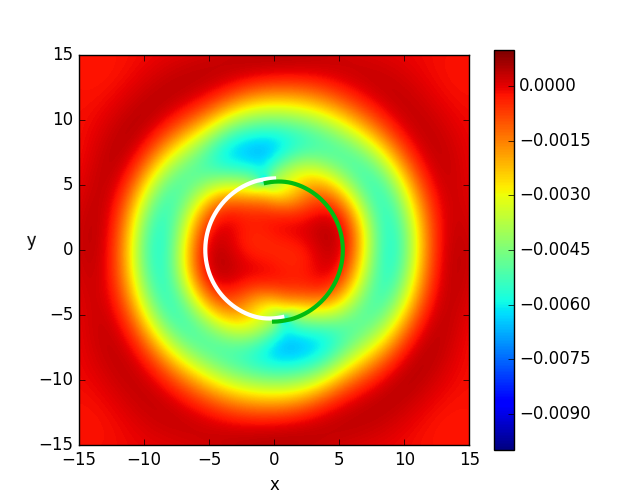}
\end{minipage}
\begin{minipage}[t]{0.3\linewidth}
\centering
\includegraphics[width=2.1in]{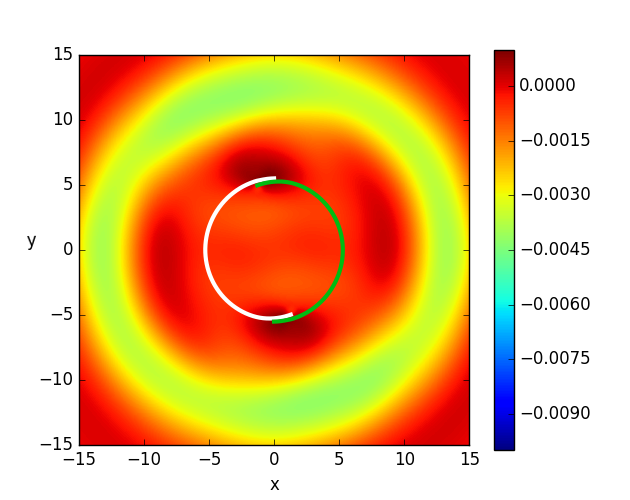}
\end{minipage}
\begin{minipage}[t]{0.3\linewidth}
\centering
\includegraphics[width=2.1in]{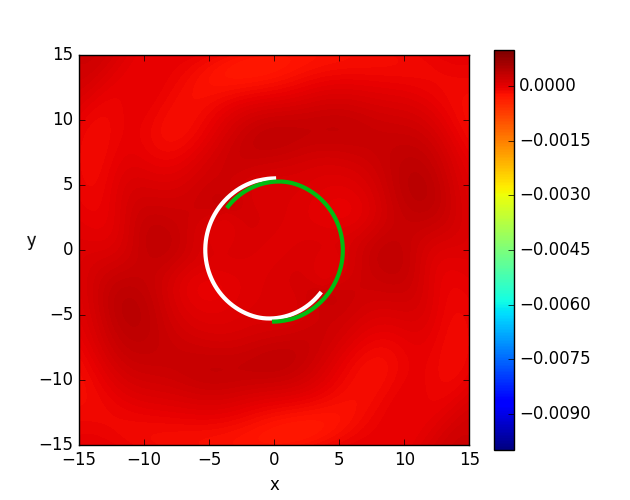}
\end{minipage}

\caption{\small  The joint evolution of the scalar field and the black hole binary system in the medium $\phi_0$ case is shown. The orbits of the two black holes are given in white and green lines. Different subplots represent different timesteps, which are chosen to be $t=111.10\mathrm{M},~116.15\mathrm{M},~121.20\mathrm{M},~126.25\mathrm{M},~131.30\mathrm{M},~
136.35\mathrm{M},~141.40\mathrm{M},~146.45\mathrm{M},~166.65\mathrm{M}$ respectively. The in-going axion shell escapes from center.}
\label{medium_phi}
\end{figure}
%\begin{figure}
%\begin{minipage}[t]{0.3\linewidth}
%\centering
%\includegraphics[width=2in]{pic213_phi/plotphi_02_00020.png}
%\end{minipage}
%\begin{minipage}[t]{0.3\linewidth}
%\centering
%\includegraphics[width=2in]{pic213_phi/plotphi_02_00120.png}
%\end{minipage}
%\begin{minipage}[t]{0.3\linewidth}
%\centering
%\includegraphics[width=2in]{pic213_phi/plotphi_02_00360.png}
%\end{minipage}
%
%\begin{minipage}[t]{0.3\linewidth}
%\centering
%\includegraphics[width=2in]{pic213_phi/plotphi_02_00480.png}
%\end{minipage}
%\begin{minipage}[t]{0.3\linewidth}
%\centering
%\includegraphics[width=2in]{pic213_phi/plotphi_02_00500.png}
%\end{minipage}
%\begin{minipage}[t]{0.3\linewidth}
%\centering
%\includegraphics[width=2in]{pic213_phi/plotphi_02_00520.png}
%\end{minipage}
%
%\begin{minipage}[t]{0.3\linewidth}
%\centering
%\includegraphics[width=2in]{pic213_phi/plotphi_02_00540.png}
%\end{minipage}
%\begin{minipage}[t]{0.3\linewidth}
%\centering
%\includegraphics[width=2in]{pic213_phi/plotphi_02_00640.png}
%\end{minipage}
%\begin{minipage}[t]{0.3\linewidth}
%\centering
%\includegraphics[width=2in]{pic213_phi/plotphi_02_01020.png}
%\end{minipage}
%\caption{fig2}
%\label{fig:side:b}
%\end{figure}
\begin{figure}
\begin{minipage}[d]{0.48\linewidth}
\centering
\includegraphics[width=3.1in]{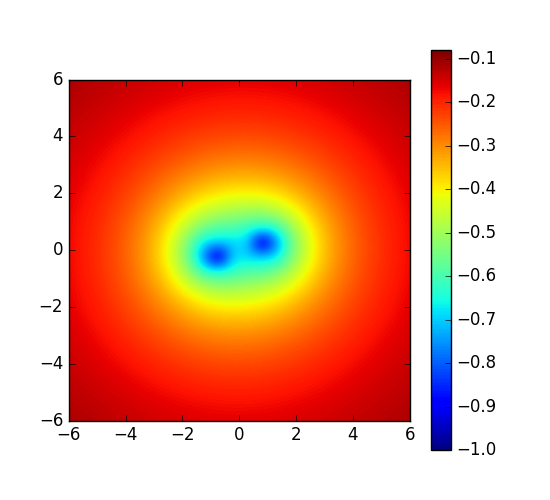}
\end{minipage}
\begin{minipage}[d]{0.4\linewidth}
\centering
\begin{minipage}[t]{0.3\linewidth}
\centering
\includegraphics[width=0.8in]{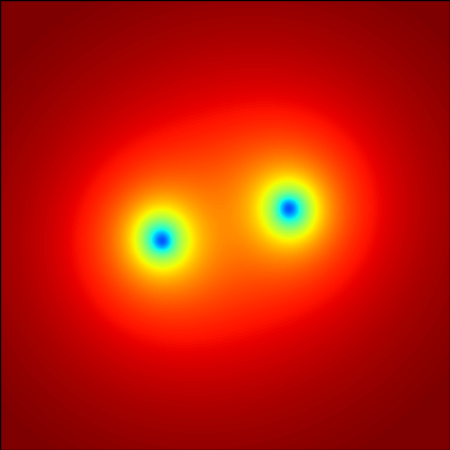}
\end{minipage}
\begin{minipage}[t]{0.3\linewidth}
\centering
\includegraphics[width=0.8in]{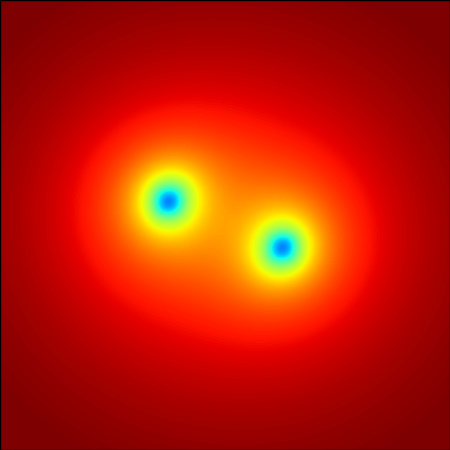}
\end{minipage}
\begin{minipage}[t]{0.3\linewidth}
\centering
\includegraphics[width=0.8in]{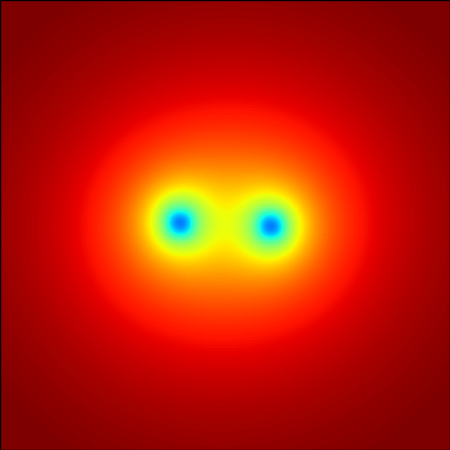}
\end{minipage}

\begin{minipage}[t]{0.3\linewidth}
\centering
\includegraphics[width=0.8in]{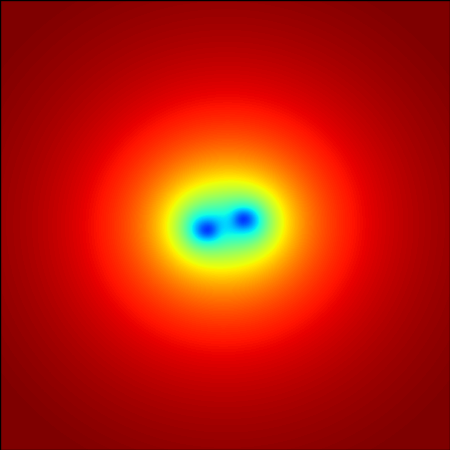}
\end{minipage}
\begin{minipage}[t]{0.3\linewidth}
\centering
\includegraphics[width=0.8in]{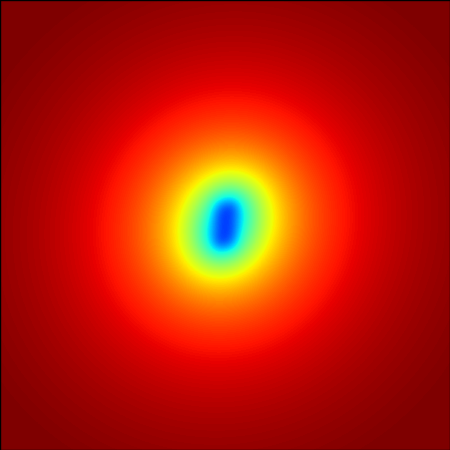}
\end{minipage}
\begin{minipage}[t]{0.3\linewidth}
\centering
\includegraphics[width=0.8in]{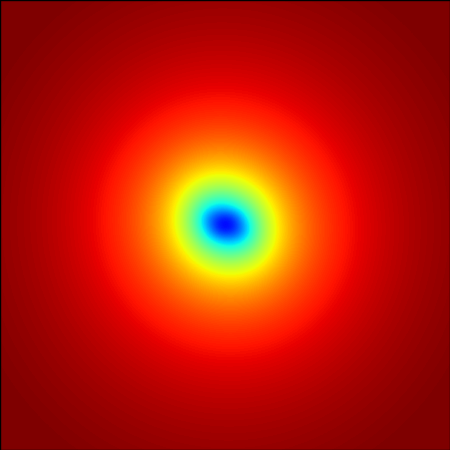}
\end{minipage}

\begin{minipage}[t]{0.3\linewidth}
\centering
\includegraphics[width=0.8in]{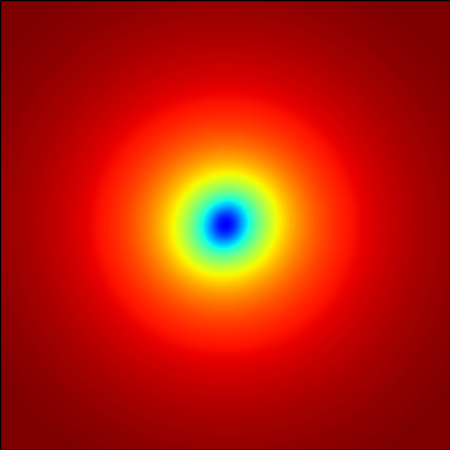}
\end{minipage}
\begin{minipage}[t]{0.3\linewidth}
\centering
\includegraphics[width=0.8in]{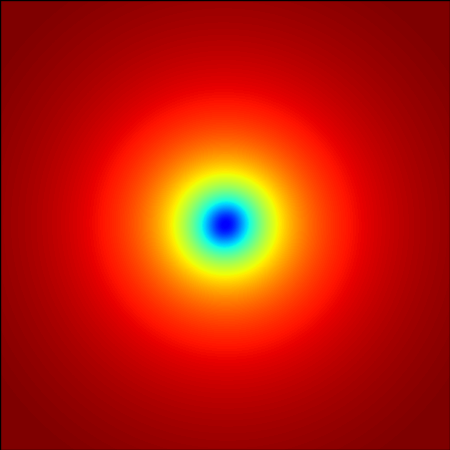}
\end{minipage}
\begin{minipage}[t]{0.3\linewidth}
\centering
\includegraphics[width=0.8in]{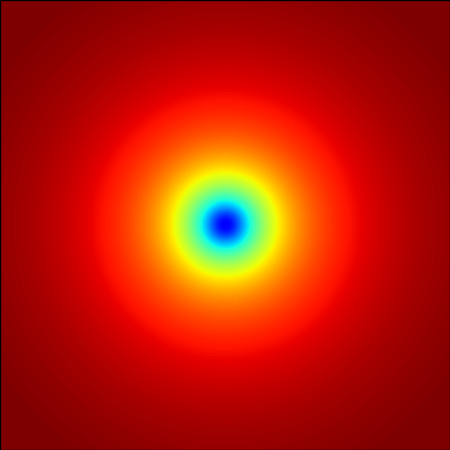}
\end{minipage}
\end{minipage}
\begin{minipage}[d]{0.1\linewidth}
\centering
\includegraphics[width=0.29in]{pic212_Lap/colorbar.png}
\end{minipage}
\caption{\small $\alpha-1$ in the medium $\phi_0$ case is shown, where $\alpha$ is the lapse function. Left panel: a detailed sketch of $\alpha-1$ at $t=1232.20\mathrm{M}$, there is no sign of a third black hole in the center. Right panel: sketch of $\alpha-1$ in different timesteps, which is $t=1105.95\mathrm{M},~1146.35\mathrm{M},~1191.80\mathrm{M},~1217.05\mathrm{M},~1222.10\mathrm{M},
~1227.15\mathrm{M},~1232.20\mathrm{M},~1242.30\mathrm{M},$ $~1262.50\mathrm{M}$ respectivily. The final black hole is much larger than the two initial black holes.}
\label{medium_lap}
\end{figure}

In this section, we will present the numerical results of the evolution of the system in the spherical symmetric medium and small $\phi_0$ cases, i.e. $\phi_0$ equals to $10^{-3}$ and $10^{-4}$ respectively.

For the medium $\phi_0$ case, the qualitative behavior of the axion field in the early time is the same with the large $\phi_0$ case, the shell splits into two parts and evolves inward and outward separately. The evolution of the ingoing shell together with the black hole orbits is shown in Fig.\ref{medium_phi}. We can see that after the ingoing shell falls to the center it escapes, propogates outward and then disappears outside the numerical boundary after some time. This behavior is different from the large $\phi_0$ case.

Since the axion shell escapes from the center, we can guess that there won't be a third black hole formed in this process. This can again be confirmed by looking at the lapse function data, which is shown in Fig.\ref{medium_lap}. As before, we show the evolution of $\alpha-1$ through different timesteps in the righthand side of Fig.\ref{medium_lap}, and in the lefthand side is a detailed sketch of the $\alpha-1$ at $t=171.70\mathrm{M}$. This time, the data shows no sign of a third black hole in the center. The final black hole is not so large as the first case, but is comparable with the initial black holes.

\begin{figure}
\begin{minipage}[t]{0.5\linewidth}
\centering
\includegraphics[width=2.7in]{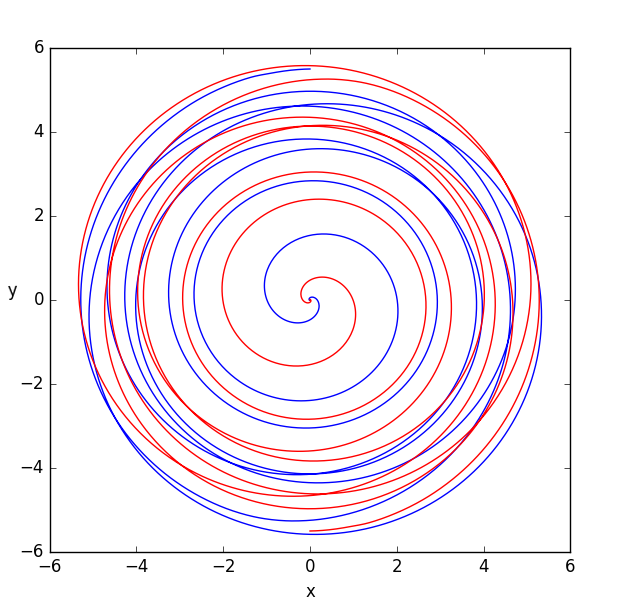}
\end{minipage}
\begin{minipage}[t]{0.5\linewidth}
\centering
\includegraphics[width=3.5in]{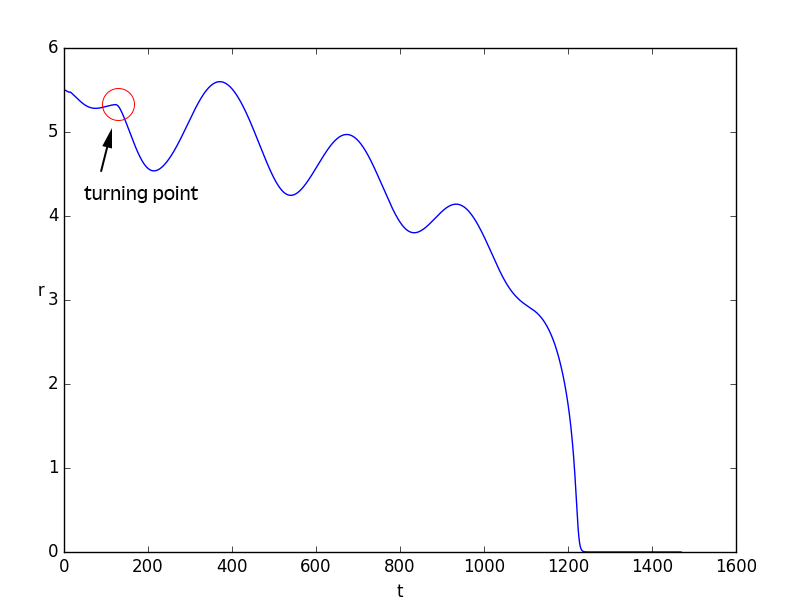}
\end{minipage}
\caption{\small Left panel: the orbits of the two black holes in the medium $\phi_0$ case. Two different colors represent two black holes. Right panel: the radial position of one of the black holes as a function of time in the medium $\phi_0$ case. The orbits gain eccentricity through the passage of the axion shell, and turn into round orbits again at the late stage of the evolution.}\label{medium_orbit}
\end{figure}

The orbits of the black holes and the radial position as a function of $t$ are shown in Fig.\ref{medium_orbit}. We can see, in this case, the orbits of the black holes are not so strongly affected by the axion shell as in the large $\phi_0$ case. But it is clear that the binary orbits developed a large amount of eccentricity (the wiggles in the right panel of Fig. \ref{medium_orbit}) through the accretion of the axion shell, and turned again into round orbits at the late stage of the evolution. A turning point can been seen in the right panel of Fig.\ref{medium_orbit} at around $t=126\mathrm{M}$. At this time, the ingoing axion shell has all gathered inside the orbits of the two black holes, as is depicted in the fourth subplot in Fig.\ref{medium_phi}. The two black holes are attracted by the axion field to fall faster to the center so that formed the turning point in the panel, and turned from semi-circular orbits to elliptic orbits. This is different from the large $\phi_0$ case that the orbits turn sharply as soon as the axion starts to collide with the BBH. So we can see that when the axion field is smaller, it interacts with the BBH mainly through gravitational attraction, not through collision.

\begin{figure}
\begin{minipage}[t]{0.5\linewidth}
\centering
\includegraphics[width=2.7in]{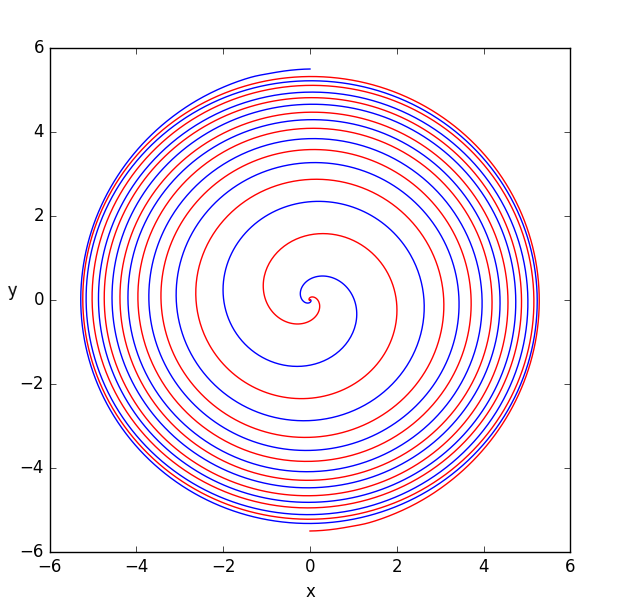}
\end{minipage}
\begin{minipage}[t]{0.5\linewidth}
\centering
\includegraphics[width=3.5in]{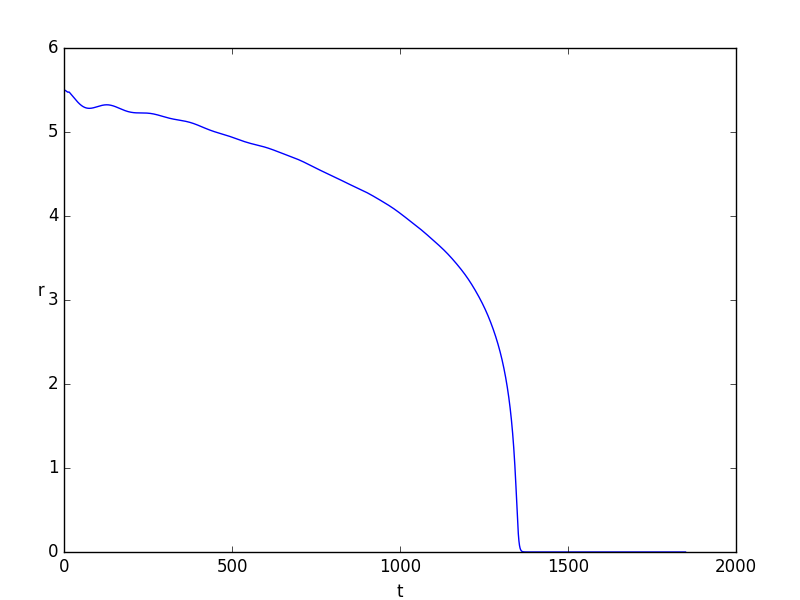}
\end{minipage}
%\begin{minipage}[t]{0.3\linewidth}
%\centering
%\includegraphics[width=2.1in]{BHr/plotBHr_GR.png}
%\end{minipage}
\caption{Left panel: the orbits of the two black holes in the small $\phi_0$ case. Two different colors represent two black holes. Right panel: the radial position of one of the black holes as a function of time in the small $\phi_0$ case. The orbits of the black holes in the small $\phi_0$ case is nearly the same as the vacuum GR case.}\label{small_orbit}
\end{figure}

For the small $\phi_0$ case, the qualitative behavior of the axion field is the same as the medium $\phi$ case: the shell splits into two parts, evolve inward and outward separately. The ingoing shell implodes through the center and then escapes to infinity. The axion field is so small that it has little impact on the BBH. The orbits of the back holes are shown in Fig.\ref{small_orbit}, and they are nearly the same as in the vacuum general relativity case.

\begin{figure}
  \centering
  \includegraphics[width=0.9\textwidth]{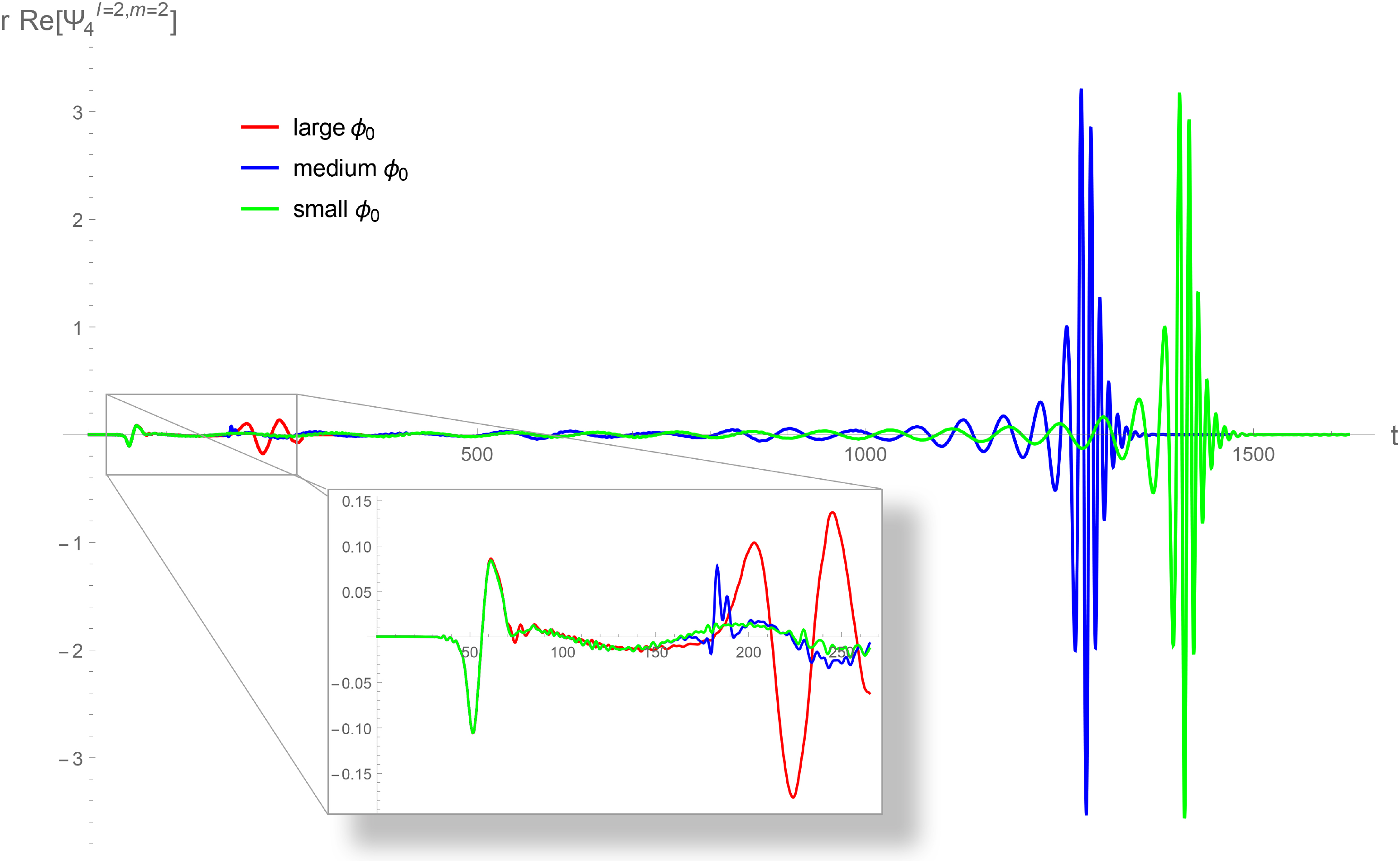}
  \caption{\small The waveform of all the three cases with spherical symmetry detected at $r=50\mathrm{M}$. For the medium $\phi_0$ case (blue curve), the initial burst produced at around $t=126\mathrm{M}$ is detected at around $t=180\mathrm{M}$ at $r=50\mathrm{M}$. For the large $\phi_0$ case (red curve), the initial burst produced at around $t=135\mathrm{M}$ is detected at around $t=220\mathrm{M}$ at $r=50\mathrm{M}$. The wiggle located at $t=50\mathrm{M}$ is from the artificial numerics.}\label{GW}
\end{figure}

The waveform of all three spherical symmetric cases is plotted in Fig.\ref{GW}. For the large $\phi_0$ case, we can see the radiated wave is greatly reduced since head-on collision suppresses the quadrupole and produces less gravitational radiation than the spherical orbits.
 For the small $\phi_0$ case, the gravitational wave emitted is nearly the same as the vacuum GR case. For the medium $\phi_0$ case, and the peak value is a little higher than the small case since elliptic orbit produces more radiation, and as a result, we can see the black holes merger faster than the small case.
In Fig.\ref{GW} we also plot the begining part of the waveforms. A small bust can be seen in the medium $\phi_0$ case. From the detection time, we find the small burst is resulted from the sudden change of the black hole orbits induced by the attraction of the axion field at around $t=126\mathrm{M}$, which has been described before in the radius-time plot of Fig.\ref{medium_orbit}, but is not resulted from the collision of the axion field and the BBH. This peak is not seen in the small $\phi_0$ case, since the axion field is so small and impact of the orbits of the black holes is too tiny to produce this kind of small burst. For the large $\phi_0$ case, the shell reaches the black hole a little bit slower than the medium case, and is followed very quickly by the collision of the two black holes, so only a burst of GW radiation caused by the collision can be seen in the waveform.

\section{Evolutions of the System in Non-Spherical Symmetric Medium $\phi_0$ case}
\label{ns}
\begin{figure}
\begin{minipage}[t]{0.3\linewidth}
\centering
\includegraphics[width=2in]{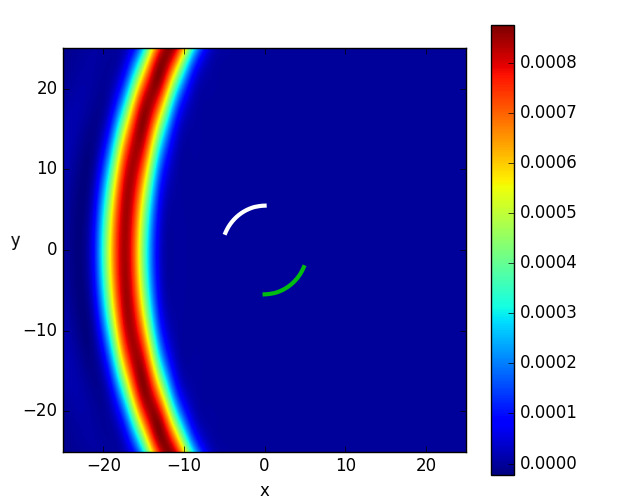}
\end{minipage}
\begin{minipage}[t]{0.3\linewidth}
\centering
\includegraphics[width=2in]{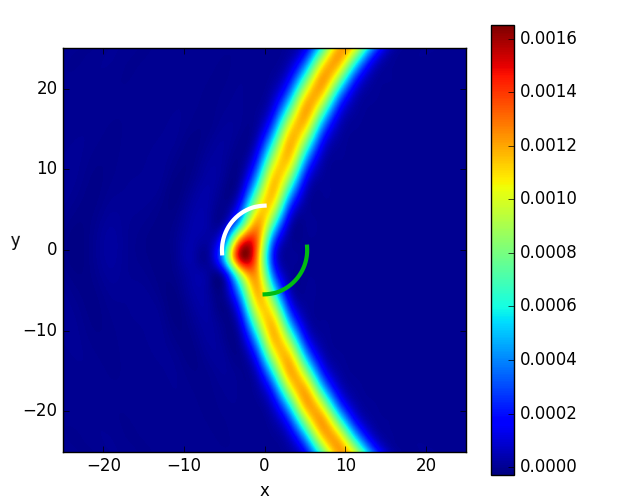}
\end{minipage}
\begin{minipage}[t]{0.3\linewidth}
\centering
\includegraphics[width=2in]{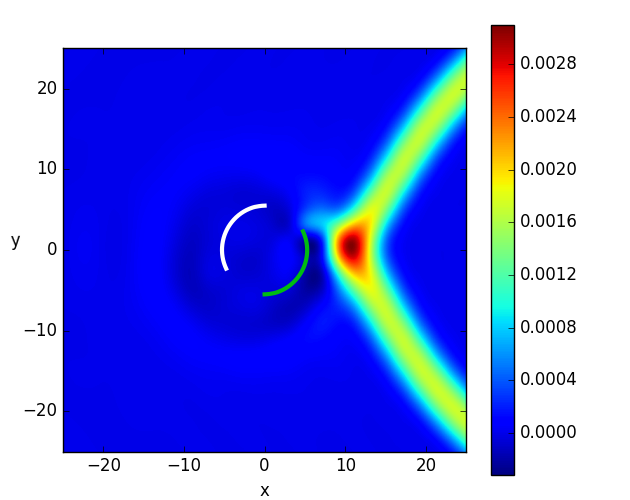}
\end{minipage}

\begin{minipage}[t]{0.3\linewidth}
\centering
\includegraphics[width=2in]{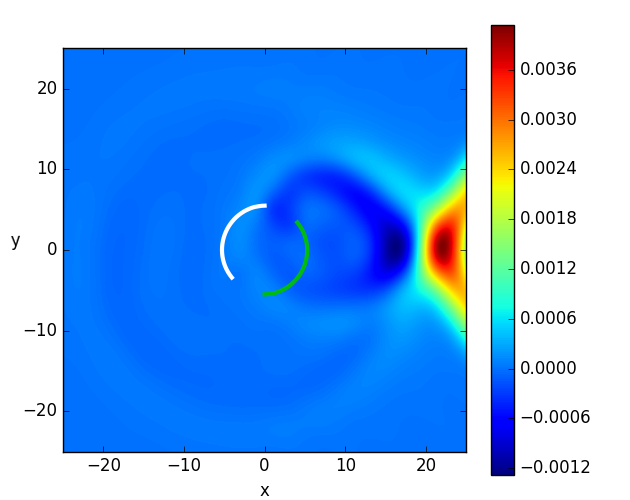}
\end{minipage}
\begin{minipage}[t]{0.3\linewidth}
\centering
\includegraphics[width=2in]{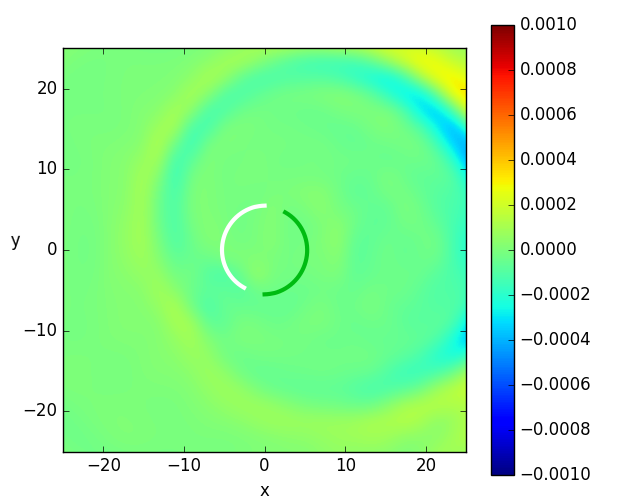}
\end{minipage}
\begin{minipage}[t]{0.3\linewidth}
\centering
\includegraphics[width=2in]{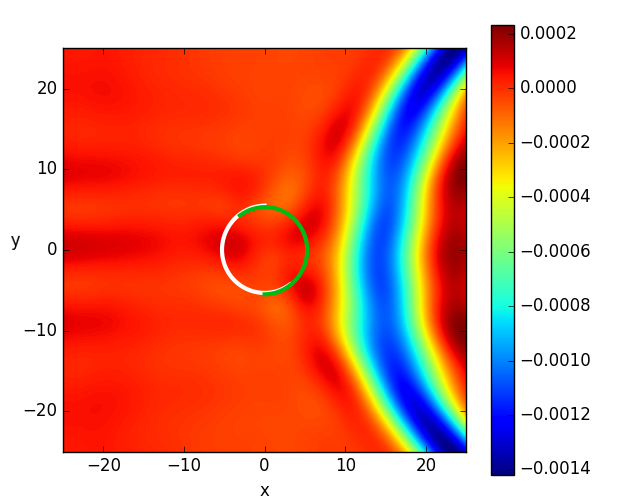}
\end{minipage}

\begin{minipage}[t]{0.3\linewidth}
\centering
\includegraphics[width=2in]{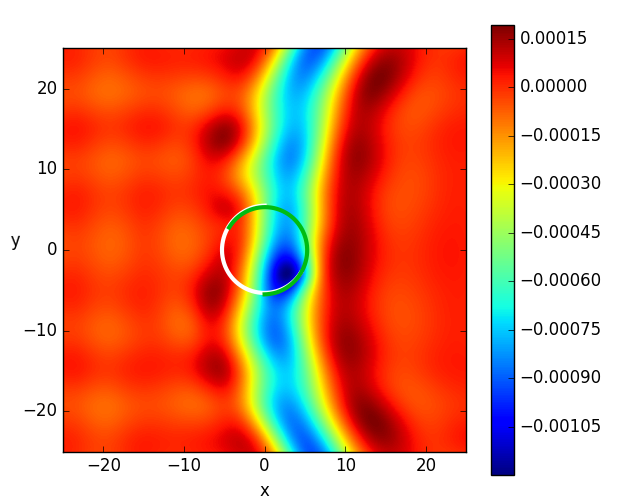}
\end{minipage}
\begin{minipage}[t]{0.3\linewidth}
\centering
\includegraphics[width=2in]{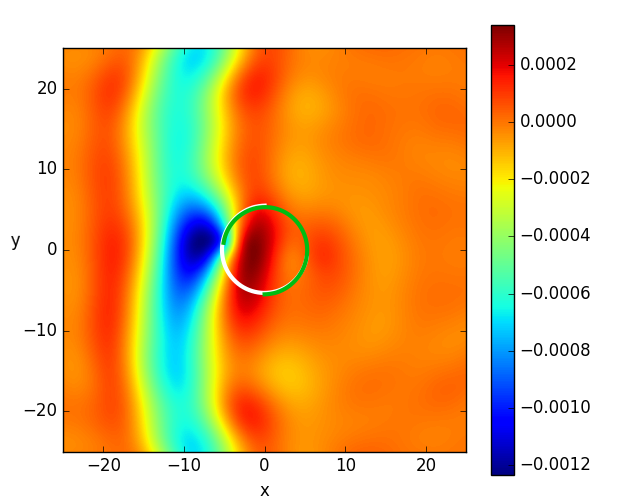}
\end{minipage}
\begin{minipage}[t]{0.3\linewidth}
\centering
\includegraphics[width=2in]{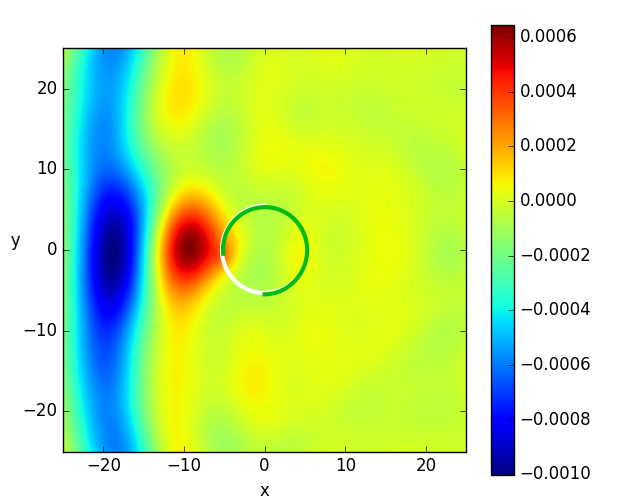}
\end{minipage}
\caption{\small The joint evolution of the scalar field and the black hole binary system in the medium $\phi_0$ case is shown. The orbits of the two black holes is given in white and green lines. Different subplots represent different timesteps, which is chosen to be $t=55.55\mathrm{M},~75.75\mathrm{M},~90.90\mathrm{M},~101.00\mathrm{M},~116.15\mathrm{M},~
161.60\mathrm{M},~176.75\mathrm{M},~191.90\mathrm{M},~202.00\mathrm{M}$ respectively. It is clear that the axion shell is dragged by the black holes every times it passes.}
\label{Nonsphe_phi}
\end{figure}
%\begin{figure}
%\begin{minipage}[t]{0.5\linewidth}
%\centering
%\includegraphics[width=2.7in]{pic214_BH/plotBH_214.png}
%\end{minipage}
%\begin{minipage}[t]{0.5\linewidth}
%\centering
%\includegraphics[width=3.5in]{BHr/plotBHr_214.png}
%\end{minipage}

\begin{figure}
\begin{minipage}[t]{0.5\linewidth}
\centering
\includegraphics[width=2.7in]{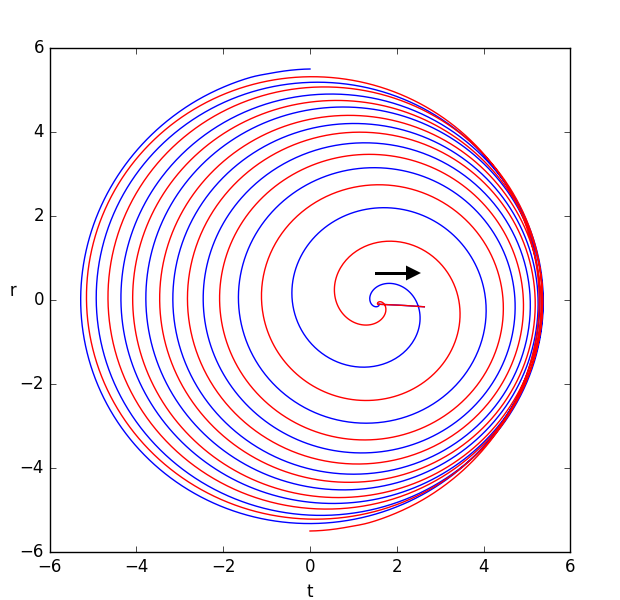}
\end{minipage}
\begin{minipage}[t]{0.5\linewidth}
\centering
\includegraphics[width=3.5in]{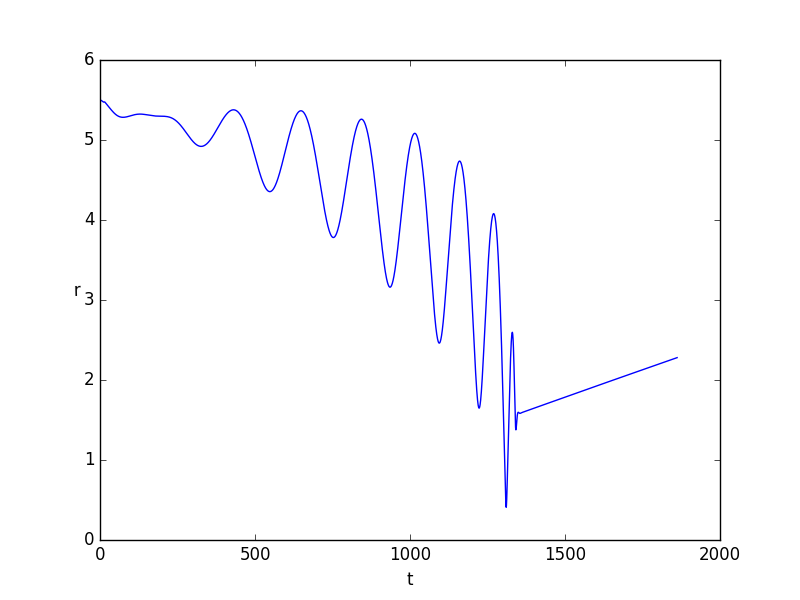}
\end{minipage}
 % \centering
  %\includegraphics[width=0.4\textwidth]{pic213NS_phi/plotBH_213.png}
  \caption{\small Left panel: The orbits of the two black holes in the medium $\phi_0$ non-spherical symmetric case. Two different color represent two different black holes. Right panel: the radial position of the black hole as a function of time in the medium $\phi_0$ non-spherical symmetric case. The black holes were attracted to the positive directions of the x-axis under the influence of the axion shell, even after the collision happened. The final velocity is $0.0014c$.}\label{Nonsphe_orbit}
\end{figure}

\begin{figure}
  \centering
  \includegraphics[width=0.9\textwidth]{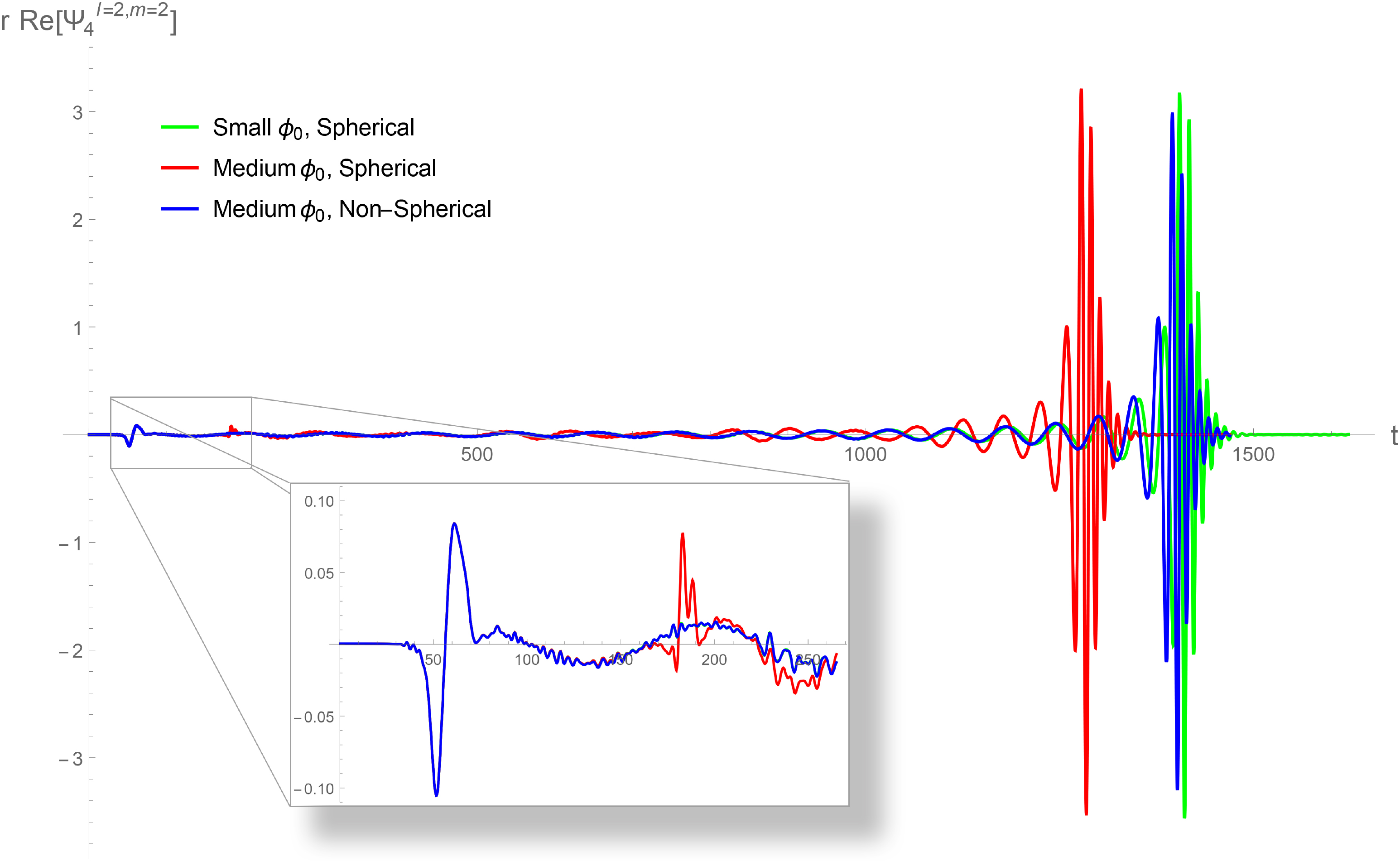}
  \caption{\small The waveform of the non-spherical case compared with the spherical symmetric with spherical symmetric case. For the spherical medium $\phi_0$ case (red), there exist a burst around $t=180\mathrm{M}$, but not in the non-spherical case (blue). The wiggle located at $t=50\mathrm{M}$ is from the artificial numerics.}\label{GW_NS}
\end{figure}

In this section we'll look at the numerical results in the non-spherical symmetric case. We'll present the evolution of the axion field, the orbits of the black holes and the cooresponding gravitational radiation emitted.

Initially, the qualitative behavior of the axion field is the same as the spherical symmetric case: the shell splits into two parts and they propagate inward and outward separately. But when the ingoing shell impacts on the binary it will be pulled back by the black holes and a sharp point will form in the axion shell. We plot the evolution of the ingoing shell together with the orbits of the black holes in Fig.\ref{Nonsphe_phi}. It is clear that the axion field is dragged back by the black hole every time it passes them. The shell implodes through center and escapes to infinity, just the same as the medium spherical symmetric case, so there is no new black hole formed.

In Fig.\ref{Nonsphe_orbit} we plot the orbits of the two black holes. we can see that, under the influence of the axion shell, the two black holes keep moving to the positive direction of the x-axis, even after the collision. So although the axion shell passes the BBH twice in total, the net impact of the axion field on the BBH is to the positive direction of the x-axis. This again confirms that at this value of the axion field, the black holes interact with the axion field mainly through gravitational attraction, but not through collisions. So when the axion shell first passes the BBH, the BBH is outside the axion shell, and the shell attracts the black holes to the positive direction of the x-axis. After the axion shell implodes through the center and passes the BBH for the second time, the BBH is again inside the shell, so there won't be much attractive force applied by the axion shell.

In Fig.\ref{GW_NS}. we plot the waveform produced in this process compared with the small $\phi_0$ and medium $\phi_0$ spherical symmetric cases. The waveform is very similar with the small $\phi_0$ case with spherical symmetry, only that the two black holes merger a little earlier. And from the subplot shown the initial part of the waveforms, we can see the small burst is not present in this non spherical symmetric case, since in this case, the net effect of the axion field on the BBH is a translation to the collapsing direction of the axion shell, and there is no suddenly fall in the initial part of the orbits as in the spherical symmetric medium $\phi_0$ case, which can be seen from the right panel of Fig.\ref{Nonsphe_orbit}. This again confirms that the small burst comes from the sudden change of orbits but not from the collisions between the axion field and the black holes.

\section{Conclusions}
\label{con}
%In summary,
In this paper, we studied how environment could modify the motion and gravitational wave signals of astrophysical black hole binaries. We considered the joint evolution of an intermediate-mass binary black hole system and a shell of axion-like scalar field around it. Both cases with and without spherical symmetry were studied.
We performed full numerical simulations and showed the numerical results concerning the evolution of the axion field, the orbits of the black holes and the gravitational wave emitted in different cases.

Firstly, when the shell of scalar field is spherical symmetric to the binary black hole mass center, we studied three cases with different $\phi_0$s, i.e. large, medium and small with $\phi_0=10^{-2}, 10^{-3}, 10^{-4}$ correspondingly. Our results show that with
increasing of the scalar field strength, the number of circular orbit of the binary black
hole is reduced. It means that the scalar field could significantly accelerate the merger
process. Once the scalar field strength exceeds some certain critical value,
the scalar field could collapse into a third black hole with its mass being larger than
the binary. In this case the orbits of the two black holes are greatly affected by the scalar field such that they are turned from quasi-circular to a head-on collision right after the pass of the scalar shell. As a result, there is almost no any quadrupole signal produced, namely the
gravitational wave is greatly suppressed. When the field strength of the scalar field is smaller than the critical value as in the medium $\phi_0$ case, the scalar field is not strong enough to form a third black hole, so the axion shell imploded through center and then escaped to infinity. The two black hole orbits develop eccentricity through the accretion of the axion field, which could accelerate the merger process compared to the semi-circular orbit. The eccentricity disappears again at the late stage of the evolution.

In the case when the axion field is non-spherical symmetric to the binary black hole system, we found that the orbits of the two black holes are constantly attracted to the mass center of the scalar field, and the existence of the scalar field slows down the merger process compared to the spherical symmetric case.

We also found that the behaviors in the initial part of the waveform can be affected by the strength of the scalar field and the symmetry property of the system. In the spherical symmetric case, when the scalar field is very small, the interaction between the scalar field and the black holes is too weak to produce any initial burst. As the strength of the scalar field gets stronger, a small burst can be found in the initial part of the waveform induced by the sudden turning of black holes orbits, which is caused by the sudden amplified attraction force to the center when the scalar field passes the two black holes and gathers inside the black hole orbits. When the strength of the scalar field gets even stronger, the collision of the black holes starts much earlier, so the initial burst is buried in the burst of radiation produced by the collision of the two black holes, and can not be seen in the waveform. In the non-spherical symmetric case, although the strength of the scalar field is the same as the medium $\phi_0$ spherical symmetric case, the initial burst does not exist in the waveform. This is because the net effect of the scalar field on the BBH is a translation along the direction of collapsing of the axion, and there is no suddenly turn to the center in the orbits as in the spherical symmetric case. This confirms that the small initial burst comes from the sudden change of orbits but not from the collisions between the scalar field and the black holes at this strength of the scalar field.

In summary, our result showed that the environmental axion-like scalar field could essentially modify the dynamics, and that it is possible to investigate the environment around binary black hole systems through the gravitational waves they emitted. The abundance and distribution of axion-like scalar field around a BBH system can be constrained through studying the overall behavior and also the initial part of the waveform.

\section*{Acknowledgements}
QY and BH are supported by the Beijing Normal University Grant under the reference No. 312232102.
BH is also partially supported by the Chinese National Youth Thousand Talents Program and the Fundamental Research Funds for the Central Universities under the reference No. 310421107. RGC and LWJ are supported in part by the National Natural Science Foundation of China Grants No.11690022, No.11375247, No.11435006, and No. 11647601, and by the Strategic Priority Research Program of CAS Grant No.XDB23030100 and by the Key Research Program of Frontier Sciences of CAS.

%%%%%%%%%%%%%%%%%%%%%%%%%%%%%%%%%%%%%%%%%%%%%%%%%%



\bibliographystyle{mnras}

%\bibliographystyle{ieeetr}
\bibliography{mnras_abh}

\begin{thebibliography}{}
\makeatletter
\relax
\def\mn@urlcharsother{\let\do\@makeother \do\$\do\&\do\#\do\^\do\_\do\%\do\~}
\def\mn@doi{\begingroup\mn@urlcharsother \@ifnextchar [ {\mn@doi@}
  {\mn@doi@[]}}
\def\mn@doi@[#1]#2{\def\@tempa{#1}\ifx\@tempa\@empty \href
  {http://dx.doi.org/#2} {doi:#2}\else \href {http://dx.doi.org/#2} {#1}\fi
  \endgroup}
\def\mn@eprint#1#2{\mn@eprint@#1:#2::\@nil}
\def\mn@eprint@arXiv#1{\href {http://arxiv.org/abs/#1} {{\tt arXiv:#1}}}
\def\mn@eprint@dblp#1{\href {http://dblp.uni-trier.de/rec/bibtex/#1.xml}
  {dblp:#1}}
\def\mn@eprint@#1:#2:#3:#4\@nil{\def\@tempa {#1}\def\@tempb {#2}\def\@tempc
  {#3}\ifx \@tempc \@empty \let \@tempc \@tempb \let \@tempb \@tempa \fi \ifx
  \@tempb \@empty \def\@tempb {arXiv}\fi \@ifundefined
  {mn@eprint@\@tempb}{\@tempb:\@tempc}{\expandafter \expandafter \csname
  mn@eprint@\@tempb\endcsname \expandafter{\@tempc}}}

\bibitem[\protect\citeauthoryear{Abbott et~al.}{Abbott
  et~al.}{2016a}]{TheLIGOScientific:2016pea}
Abbott B.~P.,  et~al., 2016a, \mn@doi [Phys. Rev.] {10.1103/PhysRevX.6.041015},
  X6, 041015

\bibitem[\protect\citeauthoryear{Abbott et~al.}{Abbott
  et~al.}{2016b}]{TheLIGOScientific:2016src}
Abbott B.~P.,  et~al., 2016b, \mn@doi [Phys. Rev. Lett.]
  {10.1103/PhysRevLett.116.221101}, 116, 221101

\bibitem[\protect\citeauthoryear{Abbott et~al.}{Abbott
  et~al.}{2016c}]{TheLIGOScientific:2016wfe}
Abbott B.~P.,  et~al., 2016c, \mn@doi [Phys. Rev. Lett.]
  {10.1103/PhysRevLett.116.241102}, 116, 241102

\bibitem[\protect\citeauthoryear{Abbott et~al.}{Abbott
  et~al.}{2016d}]{Abbott:2016nmj}
Abbott B.~P.,  et~al., 2016d, \mn@doi [Phys. Rev. Lett.]
  {10.1103/PhysRevLett.116.241103}, 116, 241103

\bibitem[\protect\citeauthoryear{Abbott et~al.}{Abbott
  et~al.}{2016e}]{Abbott:2016nhf}
Abbott B.~P.,  et~al., 2016e, \mn@doi [Astrophys. J.]
  {10.3847/2041-8205/833/1/L1}, 833, L1

\bibitem[\protect\citeauthoryear{Antonucci}{Antonucci}{1993}]{Antonucci:1993sg}
Antonucci R.,  1993, \mn@doi [Ann. Rev. Astron. Astrophys.]
  {10.1146/annurev.aa.31.090193.002353}, 31, 473

\bibitem[\protect\citeauthoryear{Baumgarte \& Shapiro}{Baumgarte \&
  Shapiro}{2003}]{Thomas:2003}
Baumgarte T.,  Shapiro S.,  2003, \mn@doi [Phys. Rep.]
  {10.1016/PhysRep.376.41}, 376, 41

\bibitem[\protect\citeauthoryear{Berti, Cardoso, Gualtieri, Horbatsch  \&
  Sperhake}{Berti et~al.}{2013}]{Berti:2013gfa}
Berti E.,  Cardoso V.,  Gualtieri L.,  Horbatsch M.,   Sperhake U.,  2013,
  \mn@doi [Phys. Rev.] {10.1103/PhysRevD.87.124020}, D87, 124020

\bibitem[\protect\citeauthoryear{{Bombaci}}{{Bombaci}}{1996}]{1996A&A...305..871B}
{Bombaci} I.,  1996, \aap, \href
  {http://adsabs.harvard.edu/abs/1996A%26A...305..871B} {305, 871}

\bibitem[\protect\citeauthoryear{Bowen \& York}{Bowen \&
  York}{1980}]{Bowen:1980yu}
Bowen J.~M.,  York Jr. J.~W.,  1980, \mn@doi [Phys. Rev.]
  {10.1103/PhysRevD.21.2047}, D21, 2047

\bibitem[\protect\citeauthoryear{Brandt \& Bruegmann}{Brandt \&
  Bruegmann}{1997}]{Brandt:1997tf}
Brandt S.,  Bruegmann B.,  1997, \mn@doi [Phys. Rev. Lett.]
  {10.1103/PhysRevLett.78.3606}, 78, 3606

\bibitem[\protect\citeauthoryear{Cao, Yo  \& Yu}{Cao et~al.}{2008}]{Cao:2008wn}
Cao Z.-j.,  Yo H.-J.,   Yu J.-P.,  2008, \mn@doi [Phys. Rev.]
  {10.1103/PhysRevD.78.124011}, D78, 124011

\bibitem[\protect\citeauthoryear{Cao, Galaviz  \& Li}{Cao
  et~al.}{2013}]{Cao:2013osa}
Cao Z.,  Galaviz P.,   Li L.-F.,  2013, \mn@doi [Phys. Rev.]
  {10.1103/PhysRevD.87.104029}, D87, 104029

\bibitem[\protect\citeauthoryear{Gebhardt, Rich  \& Ho}{Gebhardt
  et~al.}{2005}]{Gebhardt:2005cy}
Gebhardt K.,  Rich R.~M.,   Ho L.~C.,  2005, \mn@doi [Astrophys. J.]
  {10.1086/497023}, 634, 1093

\bibitem[\protect\citeauthoryear{Healy, Bode, Haas, Pazos, Laguna, Shoemaker
  \& Yunes}{Healy et~al.}{2012}]{Healy:2011ef}
Healy J.,  Bode T.,  Haas R.,  Pazos E.,  Laguna P.,  Shoemaker D.~M.,   Yunes
  N.,  2012, \mn@doi [Class. Quant. Grav.] {10.1088/0264-9381/29/23/232002},
  29, 232002

\bibitem[\protect\citeauthoryear{Hu, Barkana  \& Gruzinov}{Hu
  et~al.}{2000}]{Hu:2000ke}
Hu W.,  Barkana R.,   Gruzinov A.,  2000, \mn@doi [Phys. Rev. Lett.]
  {10.1103/PhysRevLett.85.1158}, 85, 1158

\bibitem[\protect\citeauthoryear{Hui, Ostriker, Tremaine  \& Witten}{Hui
  et~al.}{2017}]{Hui:2016ltb}
Hui L.,  Ostriker J.~P.,  Tremaine S.,   Witten E.,  2017, \mn@doi [Phys. Rev.]
  {10.1103/PhysRevD.95.043541}, D95, 043541

\bibitem[\protect\citeauthoryear{Urry \& Padovani}{Urry \&
  Padovani}{1995}]{Urry:1995mg}
Urry C.~M.,  Padovani P.,  1995, \mn@doi [Publ. Astron. Soc. Pac.]
  {10.1086/133630}, 107, 803

\makeatother
\end{thebibliography}

%\begin{thebibliography}{99}

%\end{thebibliography}

%%%%%%%%%%%%%%%%%%%%%%%%%%%%%%%%%%%%%%%%%%%%%%%%%%

%%%%%%%%%%%%%%%%% APPENDICES %%%%%%%%%%%%%%%%%%%%%

\appendix
\section{Mathematical Background and Numerical Method}
\label{appendixA}
%...
%\subsection{Mathematical backgound}
%In this subsection we will describe the mathematical background and numerical method used in the numerical calculations. In general relativity, the Hilbert-Einstein action with the axion-like scalar field described above can be written as:
%\beq
%L=\int d^3x\sqrt{-g}\left[R+\frac{1}{2}g^{\mu\nu}\partial_{\mu}\phi\partial_{\nu}\phi-V(\phi)\right]
%\eeq
%where $V(\phi)$ is the potential of the scalar field, and for a axion-like field, in natural units it can be written as:
%\beq
%V(\phi)=m^2f^2\Big(1-\mathrm{cos}(\frac{\phi}{f})\Big)
%\eeq
%We can get the Einstein equation and equations of motion for the scalar field from the above action, which is:
%\beq
%G_{\mu\nu}&=&8\pi T_{\mu\nu}\\
%\square \phi&=&\frac{dV}{d\phi}\label{EOMphi}
%\eeq
%where $G_{\mu\nu}$ is the Einstein tensor, and $T_{\mu\nu}$ is the energy-momentum tensor of the scalar field, which related to the scalar field by:
%\beq
%T_{\mu\nu}=\partial_{\mu}\phi\partial_{\nu}\phi+V
%\eeq
%For our purpose of considering the interaction of axion-like dark matter and binary black holes, we will fix the mass and decay constant of the particle to $m=10^{-21}\mathrm{eV}$ and $mf=0.5\mathrm{GeV}$

%\subsection{evolution equations}
In this appendix we will describe the mathematical background and numerical method used in the numerical calculations. We use the extended AMSS-NCKU code used by Zhoujian Cao et. al. in their previous work \cite{Cao:2008wn} and \cite{Cao:2013osa}. The code is based on the BSSN formalisom, which is a conformal-traceless ``3+1" formulation of the Einstein equations. In this formalism, the spacetime is decomposed into a set of three-dimensional spacelike slices ${\Sigma_c}$, and can be described by a three-metric $\gamma_{ij}$. The extrinsec curvature specifies how these three-surfaces embedded in the four-dimensional spacetime, and the lapse function $\alpha$ and the shift vector $\beta^i$ specify how the coordinate of the three-metric is carried along the set of hypersurfaces. Following the notations of \cite{Thomas:2003}, the metric $\gamma_{ij}$ is conformally transformed via
\beq\label{ConforMetric}
\gamma_{ij}= \psi_0^{4}\bga_{ij}, ~~\psi_0=\gamma^{-1/12}
\eeq
where $\gamma$ is the determinant of the three-metric and $\psi_0$ is the conformal factor. We have chosen that $\bga=1$. So we can now treat the conformal factor $\psi$ and the conformal transformed metric $\bga_{ij}$ as two indenpendent variables, and $\bga_{ij}$ is subjected to the constraint that it is unimodular. The same can be done to the extrinsic curvature $k_{\mu\nu}$. That is, in place of $K_{ij}$ we evolve:
\beq
K\equiv\gamma^{ij}K_{ij}, ~A_{ij}\equiv \psi_0^{-4}K_{ij}-\frac{1}{3}\bga_{ij}K
\eeq
Similar to the conformal metric, $A_{ij}$ is subjected to a constraint that $A_{ij}$ is traceless. We further introduce another evolution variable, the conformal connections:
\beq
\bg^i\equiv-\bga^{ij}_{~~,j}
\eeq
which is defined in terms of the contraction of the spatial derivative of the inverse conformal three-metric $\bga^{ij}$.

The Einstein equation and equations of motion for the scalar field can be derived from the lagrangian in eq.~\eqref{Lagrangian}, which is:
\beq\label{Einstein}
G_{\mu\nu}&=&8\pi T_{\mu\nu}\\
\square \phi&=&\frac{dV}{d\phi}\label{EOMphi}
\eeq
where $G_{\mu\nu}$ is the Einstein tensor, $V$ is the potential of the axion field given in eq.~\eqref{potential} and $T_{\mu\nu}$ is the energy-momentum tensor of the scalar field.
%\beq
%T_{\mu\nu}=\partial_{\mu}\phi\partial_{\nu}\phi+V
%\eeq
For the scalar field equation \eqref{EOMphi}, we can decompose it using the 3+1 formalism described above. Defining an auxiliary variable $\varphi\equiv\mathcal{L}_n\phi$, where $\mathcal{L}_n$ denotes the Lie derivative along the normal to the hypersurface $\Sigma_t$, the evolution equation for the scalar field $\phi$ and the Lie derivative $\varphi$ in terms of the lapse function $\alpha$ and the shift vector $\beta^i$ can be expressed as :
\beq
\partial_t\phi &=& \alpha\vp+\beta^i\partial_i\phi\\
\partial_t\vp &=& \alpha\psi_0^{-4}\[\bga^{ij}\partial_i\partial_j\phi
-(\bg^i-2\bga^{ij}\psi_0\partial_j\psi_0)\partial_i\phi\]
+\psi_0^{-4}\bga^{ij}\partial_i\alpha\partial_j\phi+\alpha\vp
\eeq
where the BSSN metric conformal transformation \eqref{ConforMetric} and the following relationships are used:
\beq
K&=&-\frac{\gamma^{ij}}{2\alpha}\frac{\partial \gamma_{ij}}{\partial t},\\
\Gamma^i&=&-\frac{1}{\sqrt{\gamma}}\partial_j(\sqrt{\gamma\gamma^{ij}})
\eeq
%In summary, we have five evolution variables, i.e., $\psi$, $\bga_{ij}$, $K$, $A_{ij}$ and $\bg$, so will need five evolution
With the 3+1 formalism, the different components of the scalar field energy-momentum tensor are given by:
\beq\label{EM}
E&:=&n_a n_b T^{ab}=\frac{1}{2}D_i\phi D^i\phi+\frac{1}{2}\vp^2+V\\
s_i&:=&-\gamma_{ia}n_bT^{ab}=-\vp D_i\phi\\
s_{ij}&:=&\gamma_{ia}\gamma_{jb}T^{ab}=D_i\phi D_j\phi-\psi_0^{4}\bga_{ij}(\frac{1}{2}D_k\phi D^k\phi-\frac{1}{2}\vp^2+V)\label{EM2}
\eeq
So Einstein equation in \eqref{Einstein} can be decomposed into the five evolution equations for the dynamical variables defined above, together with the gauge conditions, the evolution equations for the system read:
\beq
\partial_t\psi_0 &=& \beta^i\psi_{0,i}-\frac{1}{6}\alpha K+\frac{1}{6}\beta^i_{~,i}\\
\partial_t\bar{\gamma}_{ij} &=& \beta^k\bar{\gamma}_{ij,k}-2\alpha A_{ij}+2\bga_{k(i}\beta^k_{~,j)}-\frac{2}{3}\bga_{ij}\beta^k_{~,k}\\
\partial_t K &=& \beta^iK_{,i}-D^2\alpha+\alpha\[A_{ij}A^{ij}+\frac{1}{3}K^2+4\pi(\rho+s)\]\\
\partial_t A_{ij} &=& \beta^k A_{ij,k}+\psi_0^{-4}\[\alpha(R_{ij}-8\pi s_{ij})-D_iD_j\alpha\]^{TF}
+\alpha(K A_{ij}-2A_{ik}A^k_j)\nn\\&+&2A_{k(i}\beta^k_{~,j)}-\frac{2}{3}A_{ij}\beta^k_{~,k}\\
\partial_t\bg^i &=& \beta^i\bg^i_{~,j}-2A^{ij}\alpha_{,j}+2\alpha(\bg^i_{jk}A^{kj}-\frac{2}{3}\bga^{ij}K_{,j}-8\pi\bga^{ij}s_j+6A^{ij}\phi_{,j})\nn\\
&-&\bg^j\beta^i_{~,j}+\frac{2}{3}\bg^i\beta^j_{~,j}+\frac{1}{3}\bga^{ki}\beta^j_{~,jk}+\bga^{kj}\beta^i_{~,kj}\\
\partial_t\alpha &=& -2\alpha K\\
\partial_t\beta^i&=&\frac{3}{4}B^i\\
\partial_t B^i&=&\partial_t\bg^i-2 B^i%+\lambda_3\beta^j\beta^i_{~,j}-\lambda_4\beta^j\bg^i_{~,j}
\eeq
where ``TF" means the trace free part.
%\subsection{constrain equations and initial data}

We still need the constrain equations to solve for the initial data. Under a 3+1 decomposition, the momentum constraint and the Hamiltonian constrain equations are:
\beq
D_jK^j_{~i}-D_iK&=&8\pi p_i\\
R+K^2-K_{ij}K^{ij}&=&16\pi E
\eeq
where $D_j$ is the covariant derivative associated with the 3-metric $\gamma_{ij}$. $E$ and $p_i$ are the energy and momentum densities given in \eqref{EM}-\eqref{EM2}. The constraint equations can be solved with the puncture method.  Following the conformal transverse-traceless decomposition approach described above, choose a conformally flat background metric, $\bga_{ij}=\delta_{ij}$, and a maximal slice condition, $K=0$. Choose also that $\varphi=0$ initially, so that $p_i=0$. Then the constraint equations will take the form:
\beq\label{constrain1}
\partial_j A^{ij}&=&0\\
\bigtriangleup\psi_0+\frac{1}{8}A^{ij}A_{ij}\psi_0^{-7}&=&-\pi\psi_0\delta^{ij}\partial_i\phi\partial_j\phi-2\pi\psi_0^5V\label{constrain2}
\eeq
where $\bigtriangleup$ is the Laplacian operator associated with Euclidian metric. In a Cartesian coordinate system $(x^i)=(x,y,z)$, there is a non-trivial solution of eq~\eqref{constrain1} which is valid for any number of black holes \cite{Bowen:1980yu}:
\beq
A^{ij}=\Sigma\left[\frac{3}{2r_n^3}\left[x^i_nP^j_n+x^j_nP^i_n-(\delta^{ij}-\frac{x^i_nx^j_n}{r_n^2})P^n_kx^k_n\right]
+\frac{3}{2r^3_n}\big(\epsilon^{ik}_{~l}S^n_kx^l_nx^j_n+\epsilon^{jk}_{~l}S^n_kx^l_nx^i_n\big)\right]
\eeq
where $n$ is a label for punctures, $r_n=\sqrt{(x-x_n)^2+(y-y_n)^2+(z-z_n)^2}$, $\epsilon^{ik}_{~l}$ is the Levi-Civita tensor associated with the flat metric, and $P_n$ and $S_n$ are the ADM linear and angular momentum of the $n$th black hole, respectively.

The Hamiltonian constraint \eqref{constrain2} becomes an elliptic equation for the conformal factor $\psi_0$. The solution splits as a sum of a singular term and a finite correction $u$: \cite{Brandt:1997tf},
\beq
\psi_0=1+\sigma\frac{m_n}{2r_n}+u,
\eeq
after inserting the above equation to eq~\eqref{constrain2}, an elliptic equation for $u$ on $\mathbb{R}^3$ is derived, and can be solved numerically.

%\subsection{gauge conditions}
%The gauge conditions are important for the numerical simulations of dynamical spacetime, and this is especially true for the moving puncture method. The Bona-Masso type slicing gauge
%conditions [21] for the lapse function and many driver gauge conditions (e.g., the ¦£-driver) for the shift vector [22, 23] are currently the main type of gauge conditions used in the punctured black hole calculations. In this work, we will only focus on these types of gauge conditions,which can be written as
%\beq
%\partial_t\alpha &=& -2\alpha K+\lambda_1\beta^i\alpha_{,i}\\
%\partial_t\beta^i&=&\frac{3}{4}f(\alpha)B^i+\lambda_2\beta^j\beta^i_{~,j}\\
%\partial_t B^i&=&\partial_t\bg^i-\eta B^i+\lambda_3\beta^j\beta^i_{~,j}-\lambda_4\beta^j\bg^i_{~,j}\nn
%\eeq
%where $\eta$ and the and the four $\lambda$s are the parameters to be chosen.
%(the parameters chosen...)
\section{From Natural Units to Geometric Units}
\label{appendixB}
In this section, we will convert the above formulas concerning the axion field from the nature units, which is usually widely used in field theory and particle physics, to geometric units, which is more convenient for numerical calculations. The three constants concerning these two units are the Plank constant $\hbar=\frac{1}{2\pi}6.626\times10^{(-34)}\mathrm{J\cdot s}$, the Newton constant $G=6.673\times10^{(-11)}\mathrm{m^3/(kg\cdot s^2)}$, and the speed of light $c=2.9979\times10^8\mathrm{m/s}$. In the natural units, the speed of light and the Plank constant are normalized to $1$, i.e. $\hbar=c=1$. Whereas in the geometric units, the speed of light and the Newton constant are normalized to $1$, $G=c=1$.

The action of the axion field in the natural units is given in eq.~(\ref{Lagrangian}).
%\beq
%L=\int d^3x\sqrt{-g}\left[\frac{1}{2}g^{\mu\nu}\partial_{\mu}\phi\partial_{\nu}\phi-m^2f^2\Big(1-\mathrm{cos}(\frac{\phi}{f})\Big)\right]
%\eeq
As mentioned above, we have chosen that $m=10^{-21}\mathrm{eV}$, and $mf=0.5 \mathrm{GeV}$. The units of the two parameters and the scalar field in the three different units systems are summerized in Table \ref{Tablemf}
\begin{table}\centering
\begin{tabular}{|c|c|c|c|}
  \hline
    & International & Natural & Geometric \\
  \hline
  f & $\mathrm{s^{-1}}$ & $\mathrm{eV}$ & $\mathrm{m^{-1}}$ \\
  \hline
  m & $\mathrm{kg}$ & $\mathrm{eV}$ & $\mathrm{m}$ \\
  \hline
  $\phi$ & $\mathrm{\sqrt{eV/m}}$ & $\mathrm{eV}$ & 1 \\
  \hline
  \end{tabular}
  \caption{Different dimensions of the axion mass $m$, decay constant $f$ and the scalar field $\phi$ in three unit systems}
\label{Tablemf}
\end{table}

In order to have geometric units expressions, we will first convert the Lagrangian of the axion field to the SI system. We know that a Lagrangian has a dimension of energy in the SI system, i.e., $\mathrm{dim}L=M L^2/T^2$. In order to have every term in the Lagrangian has this energy dimension in the SI system, we should insert back proper combinations of $\hbar$ and $c$. It turns out that a factor of $c/\hbar$ should be inserted in front of the potential term of the scalar field. On the other hand,
from the expression of the potential we know that, the argument inside the cosine function should be dimensionless, so a factor of $\sqrt{\hbar/c}$ should also be inserted into the denominator of this argument. Now we have the Lagrangian in the SI system, which is:
\beq
L=\int d^3x\sqrt{-g}\left[\frac{1}{2}g^{\mu\nu}\partial_{\mu}\phi\partial_{\nu}\phi-\frac{c}{\hbar}m^2f^2[1-cos(\phi/(\sqrt{\frac{\hbar}{c}}f))]\right]
\eeq
where we have chosen:
\beq
(\sqrt{\frac{\hbar}{c}}f)_{Nat}=0.508 M_{pl},~(\frac{c}{\hbar}m^2f^2)_{Nat}=1.486\times10^{12}\mathrm{eV}^4
\eeq
and the subscript ``Nat" means natural units, and $M_{pl}=1/\sqrt{8\pi G_N}=2.4\times10^{18}\mathrm{GeV}$ is the reduced Planck mass. We can see form the Lagrangian that it's the combination of $\frac{c}{\hbar}m^2f^2$ and $\sqrt{\frac{\hbar}{c}}f$ that will enter the equations of motion. Dimensions of these two combinations are summarized in Table \ref{TableCom}. Now we need to get the values of the two combinations in the geometric units. Firstly their values in the SI system can be found by inserting back $\hbar$ and $c$:
\beq
(\sqrt{\frac{\hbar}{c}}f)_{SI}&=&(\sqrt{\frac{\hbar}{c}}f)_{Nat}\times\sqrt{\frac{1}{\hbar c}}=2.745\times10^{30}\mathrm{\sqrt{\frac{eV}{m}}}\nn\\
(\frac{c}{\hbar}m^2f^2)_{SI}&=&(\frac{c}{\hbar}m^2f^2)_{Nat}/(\hbar^3c^3)=1.934\times10^{32}\mathrm{\frac{eV}{m^3}}
\eeq
where the subscript ``SI" means SI system. Now we can change to the geometric units by dividing proper combinations of $G$ and $c$:
\beq
(\sqrt{\frac{\hbar}{c}}f)_{Geo}&=&(\sqrt{\frac{\hbar}{c}}f)_{SI}\times\sqrt{\frac{G}{c^4}}=0.100\nn\\
(\frac{c}{\hbar}m^2f^2)_{Geo}&=&(\frac{c}{\hbar}m^2f^2)_{SI}\times\frac{G}{c^4}=2.560\times10^{-20}\mathrm{M^{-2}}
\eeq
where the subscript ``Geo" means geometric units.
We will use these geometric units values in the numerical calculations.

\begin{table}\centering
\begin{tabular}{|c|c|c|c|}
  \hline
    & International & Natural & Geometric \\
  \hline
  $\frac{c}{\hbar}m^2f^2$ &$\mathrm{kg/(m\cdot s^2)}$  &$(\mathrm{eV})^4$ & $\mathrm{m}^{-2}$ \\
  \hline
  $\sqrt{\hbar/c}$ f & $ \mathrm{\sqrt{eV/m}}$ & $\mathrm{eV}$ & $1$\\
  \hline
\end{tabular}
 \caption{Different dimensions of $\frac{c}{\hbar}m^2f^2$ and $\sqrt{\frac{\hbar}{c}}f$, which are the two combinations appeared in the Lagrangian, in three different unit systems}
\label{TableCom}
\end{table}

\bsp	% typesetting comment
\label{lastpage}
\end{document}